\documentstyle[amsmath,amssymb,aps,epsf,eqsecnum,graphicx]{revtex}
\begin{document}
\title{Random unistochastic matrices}
\author
{Karol {\.Z}yczkowski$^{1,2}$, Marek Ku{\'s}$^1$, Wojciech
S{\l}omczy{\'n}ski$^3$ and Hans--J{\"u}rgen Sommers$^4$}
\address{$^1$Centrum Fizyki Teoretycznej, Polska
 Akademia Nauk, \\ Al. Lotnik{\'o}w 32/44, 02-668 Warszawa, Poland}
\address{$^2$ Instytut Fizyki im. M. Smoluchowskiego,
  Uniwersytet Jagiello{\'n}ski,  ul. Reymonta 4, 30-059 Krak{\'o}w, Poland}
\address{$^3$ Instytut Matematyki, Uniwersytet Jagiello{\'n}ski,
 ul. Reymonta 4, 30-059 Krak{\'o}w, Poland}
\address{$^4$ Fachbereich 7 Physik, Universit{\"a}t Essen,
 45117 Essen, Germany}
%\date{\today}
\date{ August 28, 2002}
\maketitle

\begin{abstract}
An ensemble of random unistochastic (orthostochastic) matrices is defined by
taking squared moduli of elements of random unitary (orthogonal) matrices
distributed according to the Haar measure on $U(N)$ (or $O(N)$,
respectively).
An ensemble of symmetric unistochastic matrices is obtained
with use of unitary symmetric matrices pertaining to the circular orthogonal
ensemble. We study the distribution of complex eigenvalues of
bistochastic, unistochastic and orthostochastic
matrices in the complex plane. We compute averages
(entropy, traces) over the ensembles of unistochastic matrices
and present inequalities concerning the entropies
of products of bistochastic matrices.
\end{abstract}

\medskip

\pacs{PACS numbers:  05.45.Mt, 02.50.Ga, 02.10.Ud} 
%%\footnote{Key words ?}
%%\footnote{MR ?}

\section{Introduction}

Consider a square matrix $B$ of size $N$ containing non-negative entries. 
It is called {\sl stochastic} if each row sums to unity ($\sum_{i=1}^N
B_{ij}=1$ for $j=1,\dots,N$) (for information about properties of such
matrices consult \cite{CZ98,FW79}). If, additionally, each of its columns sums
to unity, i.e., $\ \sum_{i=1}^N B_{ji}=1$ for $i=1,\dots,N$, it is called
{\sl bistochastic} (or {\sl doubly stochastic}).

Bistochastic matrices emerge in several physical problems. They are used in the
 theory of majorization \cite{MO79,AU82,An89,Bh97}, 
angular momentum \cite{Lo97}, in transfer problems,
investigations of the Frobenius-Perron operator,
and in characterization of completely positive maps acting in the space
of density matrices \cite{St02}. We shall denote by
$\Omega_N^S$ (resp. $\Omega_N^B$) the sets of stochastic (resp. bistochastic)
matrices of size $N$.

A matrix $B$ is called {\sl orthostochastic}, if there exists an orthogonal 
matrix $O$, such that $B_{ij}=O_{ij}^2$ for $i,j=1,\dots,N$. Analogously, a
matrix $B$ is called {\sl unistochastic} 
({\sl unitary-stochastic})\footnote{This notation is not unique: in some
mathematical papers (e.g. \cite{PT87}) unistochastic matrices are called
orthostochastic},  if there exists a unitary matrix $U$, such that
$B_{ij}=|U_{ij}|^2$ for $i,j=1,\dots,N$. Due to unitarity (orthogonality)
condition every unistochastic (orthostochastic) matrix is bistochastic. These
four sets of nonnegative matrices are related by the following inclusion:
$\Omega_N^O \subseteq \Omega_N^U \subseteq \Omega_N^B \subset \Omega_N^S$,
where $\Omega_N^U$ and $\Omega_N^O$ represent the sets of unistochastic
(orthostochastic) matrices. For $N>2$ all three inclusions are proper
\cite{MO79}.

Unistochastic matrices appear in analysis of models describing the time 
evolution in quantum graphs \cite{KS97,Ta00,PZK01,Ta01,PZ01} and in
description of non-unitary transformations of density matrices
\cite{Ni99b,Ni00}. Moreover, the theory of majorization and unistochastic
matrices plays a crucial role in recent research on local manipulations 
with pure states entanglement \cite{Ni99} or in 
characterizing the interaction costs of non--local
quantum gates \cite{HVC02}.

In this work we analyse the structure of the set of bistochastic (unistochastic) 
matrices of a fixed size and investigate the support of their spectra.
Knowledge of any constraints for the localisation of the eigenvalues of such a
matrix is of a direct physical importance, since the moduli of the largest
eigenvalues determine the rate of relaxation to the invariant state of the
corresponding system. We define the notion of entropy of
bistochastic matrices and 
prove certain inequalities comparing the initial and
the final entropy of any probability vector subjected to a Markov chain
described by an arbitrary bistochastic matrix. A related inequality
concerns the entropy of the product of two bistochastic matrices.

Moreover, we define physically motivated ensembles of random unistochastic
 matrices and analyse their properties. As usual, the term {\sl ensemble}
denotes a pair: a space and a probability measure defined on it (for example
the {\sl circular unitary ensemble} (CUE) represents the group $U(N)$ of
unitary matrices of size $N$ with the Haar measure \cite{Me91}). Since any
unitary matrix determines a unistochastic matrix, the Haar measure on the
unitary group $U(N)$ induces uniquely the measure in the space of
unistochastic matrices $\Omega_N^{U}$, which we shall denote by $\mu_U$.
Analogously, we shall put $\mu_O$ for the measure in $\Omega_N^{O}$ induced by
the Haar measure on the orthogonal group $O(N)$. In the sequel we shall use
the names {\sl unistochastic ensemble} (resp. {\sl orthostochastic ensemble})
for the pair $\{\Omega_N^{U}, \mu_U \}$ (resp. $\{\Omega_N^{O}, \mu_O \}$).

We compute certain averages with respect to these ensembles.
Related results were presented recently by Berkolaiko \cite{Be01} and 
Tanner \cite{Ta01}. In the latter paper the author defines unitary stochastic
ensembles, which have a different meaning: they consist of unitary matrices
corresponding to a given unistochastic matrix.

This paper is organised as follows. In section~II we review properties of 
stochastic matrices. In particular we analyse the support of spectra of
random bistochastic and stochastic matrices in the unit circle. In section~III
some results concerning majorization, ordering, and entropies of bistochastic
matrices are presented. In particular we prove subadditivity of entropy for
bistochastic matrices. Section~IV is devoted to the ensembles of
orthostochastic and unistochastic matrices; we investigate the support of
their spectra, compute the entropy averages, the average traces, and 
expectation values of the moduli of subleading eigenvalues. Some open
problems are listed in section~V.
Analysis of certain families of unistochastic matrices
and calculation of the averages with respect to the
unistochastic ensemble is relegated to the appendices.

\section{Stochastic and bistochastic matrices}

\subsection{General properties}

In this section we provide a short review of properties of stochastic and 
bistochastic matrices. The set of bistochastic matrices of size $N$ can be
viewed as a convex polyhedron in $R^{N^2}$. There exist $N!$ permutation
matrices of size $N$, obtained by interchanging the rows (or columns) of the
identity matrix. Due to the Birkhoff theorem, any bistochastic matrix can be
represented as a linear combination of permutation matrices. In other words
the set of bistochastic matrices is the convex hull of the set of permutation
matrices. By the Caratheodory theorem it is possible to use only $N^2-1$
permutation matrices to obtain a given bistochastic matrix as their convex
combination \cite{MO79}. Farahat and Mirsky showed that in this combination it
is sufficient to use $N^2-2N+2$ permutation matrices only, but this number
cannot be reduced any further \cite{FM60}. The dimension of the set of
bistochastic matrices is $(N-1)^2$. The volume of the polyhedron of bistochastic
matrices was computed by Chan and Robbins \cite{CR99}.

Due to the Frobenius--Perron theorem any stochastic matrix has at least one 
eigenvalue equal to one, and all others located at or inside the unit circle.
The eigenvector corresponding to the eigenvalue $1$ has all its components real 
and non--negative. 
For bistochastic matrices the corresponding eigenvector consists of $N$
components equal to $1/N$ and is called uniform.  A stochastic matrix $S$ is
called  {\sl reducible} if it is block diagonal, or if there exists a 
permutation $P$ which brings it into a block structure,
 \begin{equation}
S'=PSP^{-1} =
\left[
\begin{array}{cc}
 A_1 & 0 \\
 C  & A_2 \\
\end{array}
\right] ,
\label{reduci}
\end{equation}
where $A_i$ are square matrices of size $N_i<N$ for $i=1,2$, $N = N_1+N_2$. 
It is called {\sl decomposable} if one can find two permutation matrices $P$
and $Q$ such that $PSQ$ has the above form. Matrix $S$ is {\sl irreducible}
({\sl indecomposable}) if no such matrix $P$ (matrices $P$ and $Q$) exists
(exist) \cite{MO79,Me89}. An irreducible stochastic matrix cannot have two
linearly independent vectors with all components nonnegative. Any reducible
bistochastic matrix is {\sl completely reducible} \cite{Me89}, i.e., the
matrix $C$ in (\ref{reduci}) is equal to zero.

A stochastic matrix $S$ is called {\sl primitive} if there exists only one 
eigenvalue with modulus equal to one. If $S$ is primitive, then $S^k$ is
irreducible for all $k\ge 1$ \cite{MO79}. Note that the permutation matrices
$P$ with $\operatorname*{tr}P=0$ are irreducible, but non-primitive, since
$P^N$ equal to identity is reducible. 
For any primitive stochastic there exist a natural number $k$
 such that the power $S^k$ has all entries positive. 
The fact that all
the moduli of eigenvalues but one are smaller than unity 
implies the convergence  $\lim_{k\to\infty} B^k=B_*$.
 Here $B_*$ denotes the uniform bistochastic matrix ({\sl van
der Waerden matrix}) with all elements equal to $1/N$. Its spectrum consists
of one eigenvalue equal to one and $N-1$ others equal to zero. The matrix
$B_*$ saturates the well known van der Waerden inequality \cite{MO79}
concerning the permanent of the bistochastic matrices: ${\rm per}B \ge
N!/N^N$, and hence is sometimes called the {\sl minimal} bistochastic matrix
\cite{Eg81,Me89}.

Each bistochastic matrix of size $N$ may represent a transfer process at an
oriented graph consisting of $N$ nodes. If a graph is disjoint or consists of
a  Hamilton cycle (which represents a permutation of all $N$ elements), 
the bistochastic matrix is not primitive, and the modulus of the subleading
eigenvalue (the second largest) is equal to unity.

If matrices $A$ and $B$ are bistochastic, its product $C=AB$ is also 
bistochastic. However, the set of bistochastic matrices does not form a
group, since in general the inverse matrix $A^{-1}$ is not bistochastic (if it
exists). For any permutation matrix $P$ its inverse $P^{-1}=P^T$ is
bistochastic and the eigenvalues of $P$ and $P^{-1}$ are equal and belong to
the unit circle.

\subsection{Spectra of stochastic matrices}

A stochastic matrix contains only non-negative entries and due to the 
Frobenius--Perron theorem its largest eigenvalue is real. This leading
eigenvalue is equal to unity, since its spectral radius is bounded by
the largest and the smallest sum of its rows, all of which are equal to $1$.
In the simplest case of permutation matrices the spectrum consists of some
roots of unity. The eigenvalues of permutation matrices consisting of only one
cycle of length $k$ are exactly the $k$-th roots of unity.

Upper bounds for the size of the other eigenvalues are given in \cite{Sc76}.
Let $M$ denote the largest element of a stochastic matrix and $m$ the smallest. 
Then the radius $r_2$ of a subleading eigenvalue satisfies 
\begin{equation}
r_2 \le \frac{M - m}{M+m} \quad .
\label{boundmm}
\end{equation}

From this bound it follows that all subleading eigenvalues of the
 van der Waerden uniform matrix $B_*$ vanish. Another simple bound of this
kind for a matrix of size $N$ reads \begin{equation}
r_2 \le {\rm min}\{ NM-1, 1-Nm \} \quad .
\label{boundm2}
\end{equation}

For any stochastic matrix the characteristic polynomial is real, so 
we may expect a clustering of the eigenvalues of random stochastic matrices
at the real line. This issue is related to the result of Kac, who showed 
that the number of real roots of a polynomial of order $N$ with random real 
coefficients scales asymptotically like $\ln N$ \cite{Ka59}. 
The spectrum of a stochastic matrix is
symmetric with respect to the real axis.  Thus for $N=2$ all eigenvalues are
real and the support of the spectrum of the set of stochastic matrices reduces
to the interval $[-1,1]$.

%\vskip -2.0cm
\begin{figure} [htbp]
%\vskip -1.0cm
\begin{center}
\
\includegraphics[width=10.0cm,angle=0]{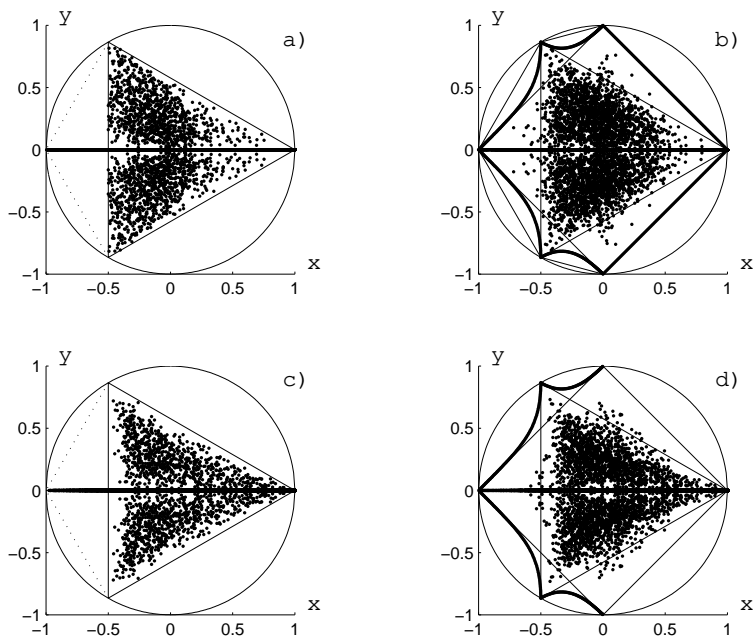}
\vskip 0.5cm
\caption{
Eigenvalues of $3000$ random stochastic matrices of size
a) N=3, b) N=4. Panels
c) and d)  show the spectra of bistochastic matrices of size $3$ and $4$.
Thick solid lines represent the bounds (\ref{arcs}) of Karpelevich.}
\label{fig1}
\end{center}
\end{figure}
%\vskip -1.0cm

\medskip

Let $Z_k$ denote a regular polygon (with its interior) centred at $0$ with one of its $k$ 
corners at $1$. The corners are the roots of unity of order $k$. Let ${\bar
E}_N$ represents the convex hull of $E_N=\bigcup_{k=2}^N Z_k$, or, in other
words, the polygon constructed of all $k$-th roots of unity, with
$k=2,\dots,N$. It is not difficult to show that the support $\Sigma_n^S$ of
the spectrum of a stochastic matrix of order $N$ is contained in ${\bar E}_n$
- see e.g. a concise proof of Schaefer \cite{Sc76}, p.15 (originally
formulated for bistochastic matrices). The $k$-th roots of the unity -- the
corners of the regular $k$--polygon, represent eigenvalues of non-trivial
permutation matrices of size $k$. For $N=3$ this polygon becomes a deltoid
(dotted lines in Fig.~1b), while for $N=4$ a non-regular hexagon.
 However, this set
is larger than required: as shown by Dimitriew and Dynkin \cite{DD46} for
small $N$ and later generalised and improved by Karpelevich \cite{Ka51} for an
 arbitrary matrix size, the support $\Sigma_N^S$ of the spectrum  of
stochastic matrices  forms, in general, a set which is not convex. For a
recent simplified proof of these statements consult papers by Djokovi{\v c}
\cite{Dj90} and Ito \cite{It97}. For instance, the support $\Sigma_3^S$
consists of a horizontal interval and the equilateral triangle (see Fig.~1a),
while for $\Sigma_4^S$ four sides of the hexagon should be replaced by the
arcs, interpolating between the roots of the unity, and given by the solutions
$\lambda =\left| \lambda \left( t\right) \right| e^{i\phi \left( t\right) }$
with $t\in \left[ 0,1\right]$ of 

\begin{eqnarray}
\lambda^4+(t-1)\lambda -t=0
\quad \text {for which} \quad |\phi(t)|\in [\pi/2,2\pi/3], \nonumber
\end{eqnarray}
and
\begin{eqnarray}
\lambda^4-2t\lambda^2+(2t-t^2-1)\lambda(t-1)+t^2=0 \quad
\text {for which} \quad |\phi(t)|\in [2\pi/3,\pi]. 
\label{arcs} 
\end{eqnarray}

\subsection{Spectra of bistochastic matrices}

The {\sl spectral gap} of a stochastic matrix is defined as $1-r_2$, where $r_2$ denotes 
the modulus of the subleading eigenvalue \cite{Be01} (note that in
\cite{Ta01} the gap is defined as $-\ln r_2$). This quantity is relevant for
several applications since it determines the speed of the relaxation to the
equilibrium of the dynamical system for which $B$ is the transition matrix.
Analysing the spectrum of a bistochastic matrix it is also interesting to
study the distance of the closest eigenvalue to unity. Since for any
bistochastic matrix $1$ is its simple eigenvalue if and only if the matrix is
irreducible, some information on the spectrum may be obtained by introducing a
measure of irreducibility. Such a strategy was pursued by Fiedler \cite{Fi95},
who defined for any bistochastic matrix $B$ the following quantities 

\begin{equation}
\mu(B):= {\rm min}_A \sum_{i\in A} \sum_{j\in {\bar A}} B_{ij}
\quad \quad {\rm and} \quad \quad
\nu(B):= {\rm min}_A \sum_{i\in A} \sum_{j\in {\bar A}} \frac{B_{ij}}{n(N-n)},
\label{munu}
\end{equation}
where $A$ is a proper subset of indices $I=\{1,2,\ldots,N \}$ containing $n$ elements, 
$1\le n < N$, and $\bar A$ denotes its complement such that $A\cup {\bar
A}=I$. Observe that $\mu$ measures the minimal total weight of the
`non-diagonal' block of the matrix, which equals to zero for reducible
matrices, while $\nu$ is averaged over the number of $n(N-n)$ elements, which
form such a block. Fiedler \cite{Fi95} used these quantities to establish the
bounds for the subleading eigenvalue ${\lambda}_2$
 \begin{equation} |1-\lambda_2|
\ge 2 \mu(B) (1-\cos\frac{\pi}{N}) \quad \quad  {\rm and} \quad \quad
|1-\lambda_2| \ge 2\nu (B).
 \label{munures} 
\end{equation}

Since $\Omega_N^B \subseteq \Omega_N^S$, the supports of the spectra
 fulfil $\Sigma_N^B \subseteq \Sigma_N^S$. Moreover, our numerical analysis
suggests that for $N\ge 4$ this inclusion is proper. In fact, they do not
contradict an appealing conjecture \cite{FW79}, that the support $\Sigma_N^B$
of the spectra of bistochastic matrices is equal to the set theoretical sum
$E_N$ of regular $k-$polygons $Z_k$ ($k=2,\ldots,N$), whose points are the
consecutive roots of the unity. Numerical results obtained for random matrices
chosen according to the uniform measure in the $(N-1)^2-$dimensional  space of
bistochastic matrices are shown in Fig.~1c~and~1d. 

It is not difficult to show that these polygons are indeed contained in the 
set $\Sigma_N^B$, i.e., for each $\lambda \in \bigcup_{k=2}^N Z_k$ there
exists an $N \times N$ bistochastic matrix $B$ such that $\lambda$ belongs to
its spectrum, $\operatorname*{Sp}(B)$. First note that the support $\Sigma_{N-1}^B$ is
included in $\Sigma_N^B$. Hence, it is enough to show that $Z_N \subset
\Sigma_N^B$. Let us start from the following simple observation: if $A,B \in
\Omega_N^B$ commute, then the corresponding eigenspaces are equal, and if
$\lambda \in \operatorname*{Sp}(A), \mu \in \operatorname*{Sp}(B)$ are the
corresponding eigenvalues, then $x\lambda+(1-x)\mu \in
\operatorname*{Sp}(xA+(1-x)B)$ for each $x \in [0,1]$. 

In the sequel $P_{(i^1_1 \dots i^1_{k_1}) \dots (i^m_1 \dots i^m_{k_m})}$ will denote
 the matrix, which corresponds to the permutation consisting of $m$ cycles:
$(i^1_1 \to \dots \to i^1_{k_1}), \dots, (i^m_1 \to \dots \to i^m_{k_m})$, where 
$k_1 + \dots + k_m = N$. Let $w_N$ be a vector whose coordinates are the
consecutive $N$th-roots of the unity, i.e., $w_N=\left( 1,\exp (2\pi
i/N\right) ,\ldots ,\exp \left( 2\pi i\left( N-1\right) /N\right)$. Let
$k=0,\ldots ,N-1$. Then matrices $P_{\left( 1\ldots N\right) }^{N-k}$ and
$P_{\left( 1\ldots N\right) }^{N-\left( k+1\right) }$ commutes, and $P_{\left(
1\ldots N\right) }^{N-k}w_{N}=\exp (2\pi ki/N) w_{N}$, $P_{\left( 1\ldots
N\right) }^{N-\left( k+1\right) }w_{N}=\exp (2\pi \left( k+1\right) i/N)
w_{N}$. Hence the edges joining the complex numbers: $\exp (2\pi ki/N)$ and
$\exp (2\pi \left( k+1\right) i/N)$ for $k=0,\ldots ,N-1$ are contained in
$\Sigma_N^B$, and so the whole boundary of the polygon $Z_N$. Furthermore,
taking linear combinations of a bistochastic matrix with an eigenvalue at the
edge of the polygon $Z_N$ and the identity matrix ${\mathbb I}_N$ we may
generate lines in $\Sigma_N^B$ joining this boundary with point $1$. It
follows therefore, that the entire inner part of each polygon constitutes a
part of the support of the spectrum of the set of bistochastic matrices.

Let us now analyse in details the case $N=3$. $P_{(123)}$ denotes the $3 \times 3$
 matrix representing the permutation $1\to 2 \to 3 \to 1$. Its third power is
equal to identity, $P_{(123)}^3=P_{(1)(2)(3)}={\mathbb I}_3$, while
$P_{(123)}^2=P_{(123)}^{-1}=P_{(132)}$. Two edges of the equilateral triangle
joined in unity are generated by the spectra of $xP_{(123)}+(1-x){\mathbb
I}_3$ - see Fig.~2b. The third vertical edge is obtained from spectra of
another interpolating family $xP_{(123)}+(1-x)P_{(132)}$ - see Fig.~2a. The
boundary of $\Sigma_3^B$ is obtained from spectra of the members of the convex
polyhedron of the bistochastic matrices of size $3$. On the other hand, the
permutations $P_{(123)}$ and $P_{(12)(3)}$ do not commute, and the spectra of
their linear combination form a curve inside $\Sigma_3^B$ - see Fig.~2c. To
show that there are no points in the set $\Sigma_N^B$ outside $E_3$ note that
$Z_2 \cup Z_3 = E_3 \subseteq \Sigma_3^B \subseteq \Sigma_3^S = E_3$.

Let us move to the case $N=4$. Two pairs of sides of the square forming $Z_4$ 
constructed of commuting permutation matrices are shown in Fig.~2d and
Fig.~2e, while Fig.~2f presents the spectra of a combination of noncommuting
permutation matrices, which interpolate between $3$ and $4$--permutations.
This illustrates the fact that $E_4$ is contained in the set $\Sigma_4^B$, but
we have not succeeded in proving that both sets are equal.

Analysis of the support of the spectra of stochastic and bistochastic matrices 
can be thus summarised by
\begin{equation}
\bigcup_{k=2}^N Z_k = E_N \subseteq \Sigma_N^B \subseteq \Sigma_N^S
={\tilde E}_N \subset {\bar E}_N,
\label{sigmabisto}
\end{equation}
where ${\tilde E}_N$ is a concave hull of the set--theoretical sum $E_N$
of the regular polygons $Z_k$ supplemented by the area bounded by the Karpelevich's 
interpolation curves \cite{Ka51,Dj90,It97}, whereas ${\bar E}_N$ is the
closed convex hull of $E_N$.

%\vskip -1.0cm
\begin{figure} [htbp]
%\vskip -1.0cm
   \begin{center}
\
 \includegraphics[width=12.0cm,angle=0]{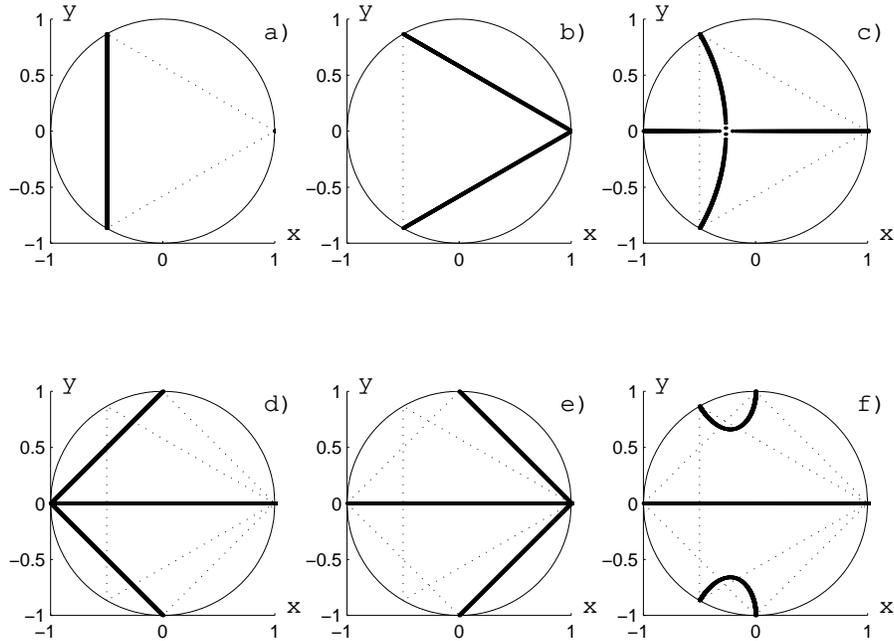}
\vskip 0.5cm
\caption{
 Eigenvalues of families of bistochastic matrices, which produce the boundary of the set $\Sigma_N^B$. They are constructed of linear combinations, 
$xP_a+(1-x)P_b$, of permutation matrices of size $N=3$:
 a) $(P_{(123)}$ and  $P_{(132)})$,
 b) $(P_{(123)}$ and  $I_3$),
 c)~$(P_{(12)(3)}$ and $I_2)$,
 and of size $N=4$:
 d) $(P_{(1432)}$ and $P_{(13)(24)})$,
 e) $(P_{(1432)}$ and $I_4) $,
 f) $(P_{(1342)}$ and $P_{(132)(4)})$.
Panels a), b), d), and f) show spectra of combinations of commuting, 
and c) and f) of noncommuting matrices.}
 \label{fig2}
\end{center}
 \end{figure}
%\vskip -2.0cm

\medskip

\section{Majorization and entropy of bistochastic matrices}

\subsection{Majorization}

Consider two vectors $\vec{x}$ and $\vec{y}$, consisting of $N$ non-negative components each. Let us assume they are normalised in a sense that $\sum_{i=1}^{N}x_{i}=\sum_{i=1}^{N}y_{i}=1$. The theory of {\sl majorization} introduces a partial order in the set of such vectors \cite{MO79}. We say that $\vec{x}$ {\sl is majorized} by $\vec{y}$, written ${\vec{x}} \prec {\vec{y}}$, if
\begin{equation}
\sum_{i=1}^{k}x_{i}\leq \sum_{i=1}^{k}y_{i}  \label{major}
\end{equation}
for every $k=1,2,\dots ,N$, where we ordered the components of each vector in the decreasing order, $x_{1}\geq \ldots \geq x_{N}$ and $y_{1}\geq \ldots \geq y_{N}$. Vaguely speaking, the vector $\vec{x}$ is more `mixed' than the vector, $\vec{y}$.

A following theorem by Hardy, Littlewood, and Polya applies the bistochastic
matrices to the theory of majorization \cite{MO79}.

{\bf Theorem 1. (HLP)} {\sl For any vectors $\vec x$ and $\vec y$, with sum of
their components normalised to unity, ${\vec x} \prec {\vec y}$ if and only if}
\begin{equation}
{\vec x}=B{\vec y} \quad {\rm for \quad some
\quad bistochastic \quad matrix} \quad B.
 \label{bimaj}
\end{equation}
It was later shown by Horn \cite{Ho54} that in the above theorem the word `bistochastic'
 can be replaced by `orthostochastic'. In general, the orthostochastic matrix
$B$ satisfying the relation ${\vec x}=B{\vec y}$ is not unique.

The functions $f$, which preserve the majorization order: 
\begin{equation}
{\vec x} \prec {\vec y} \quad {\rm implies } \quad f(\vec x) \le f(\vec y).
\label{shur}
\end{equation}
are called {\sl Schur convex} \cite{MO79}. Examples of Schur convex functions include $h_q({\vec x})=\sum_{i=1}^N x_i^q$ for any $q\ge1$, and $h({\vec x})=\sum_{i=1}^N x_i \ln x_i$.

The degree of mixing of the vector $\vec x$ can be characterised by its Shannon entropy $H$ or
 the generalised R\'{e}nyi entropies $H_q$ ($q \ge 1$):

\begin{eqnarray}
H ({\vec x}) = -\sum_{i=1}^N x_i \ln x_i , \nonumber \\
H_q ({\vec x}) = {\frac{1 }{1-q}} \ln \Bigl( \sum_{i=1}^N x_i^q \Bigr).
\label{renyi}
\end{eqnarray}
In the limiting case one obtains $\lim_{q\to 1} H_q({\vec x})= H({\vec x})$. 
If $\vec x$ represents the non-negative eigenvalues of a density matrix, $H$
is called the von Neumann entropy. Let ${\vec x}=B{\vec y}$. Due to the
Schur-convexity we have $H_q({\vec x}) \ge H_q({\vec y})$, since  we changed
the sign and reversed the direction of inequality. An interesting application
of the theory of majorization in the space of density matrices representing
mixed quantum states is recently provided by Nielsen \cite{Ni99b}. Consider a
mixed state $\rho$ with the spectrum consisted of non-negative eigenvalues
$\lambda_{1} \geq \ldots \geq \lambda_{N}$, which sum to unity. This state may
be written as a mixture of $N$ pure states,
$\rho=\sum_{i=1}^N p_i |\psi_i\rangle \langle \psi_i|$ if, and only if, ${\vec
p} \prec {\vec \lambda} = (\lambda_1,\ldots,\lambda_N)$, where ${\vec p} =
\left(p_1,\ldots,p_k \right)$ (if both vectors have different length, the
shorter is extended by a sufficient number of extra components equal to zero).
This statement is not true if the pure states states 
$|\psi_i\rangle $ are required to be distinct \cite{BE02}.

Consider now the non-unitary dynamics of the density operators given by 
$\rho\to \rho'=\sum_{j=1}^{L} q_j U_j \rho U_j^{\dagger}$. This process,
called {\sl random external fields} \cite{AL87}, is described by $L$
unitary operations $U_j$ ($j=1,\dots,L$) and non-negative probabilities
satisfying $\sum_{j=1}^L q_j=1$. Denoting respective spectra, ordered
decreasingly, by $\vec \lambda$ and ${\vec \lambda}'$ one can find
unistochastic matrix $B$ such that ${\vec \lambda}'=B{\vec \lambda}$, so due
to the HLP theorem we have $\vec{\lambda}' \prec {\vec \lambda}$
\cite{Ul71,Ni00}. Therefore, after each iteration the mixed state becomes more
mixed and its von Neumann entropy is non-decreasing.

\subsection{Preorder in the space of bistochastic matrices}

A bistochastic matrix $B$ acting on a probability vector $\vec{x}$ makes it
 more mixed and increases its entropy. To settle which bistochastic matrices
have stronger mixing properties one may introduce a relation (preordering) in
the space of bistochastic matrices \cite{Sh52} writing

\begin{equation}
B_{1}\prec B_{2}\quad\text{iff}\quad B_{1}=BB_{2} \text{   for some bistochastic
matrix }B \text{ .}
\label{maj3}%
\end{equation}
We have already distinguished some bistochastic matrices: permutation 
matrices $P$ with only $N$ non-zero elements, and the uniform van der Waerden
matrix $B_{\ast}$, with all its $N^{2}$ elements equal to $1/N$.
 For an arbitrary bistochastic matrix $B$ and for all permutation matrices
$P$ we have $B_{\ast}=B_{\ast}B$ and $B=\left( BP^{-1}\right)  P$, and hence
$B_{\ast}\prec B\prec P$. The relation $B_{1}\prec B_{2}$ implies
$B_{1}\vec{y}\prec B_{2}\vec{y}$ for every probability vector $\vec{y}$, but
the converse is not true in general (see \cite{Sh54} and \cite{Sc58}).

\subsection{Entropy of bistochastic matrices }

To measure mixing properties of a bistochastic matrix $B$ of size $N$ 
one may consider its entropy. We define \textsl{Shannon entropy} of $B$ as
the mean entropy of its columns (rows), which is equivalent to

\begin{equation}
H(B):=-\frac{1}{N}\sum_{i=1}^{N} \sum_{j=1}^{N}B_{ij}\ln B_{ij} \text{ .}
\label{sha}
\end{equation}
As usual in the definitions of entropy we set $0\ln0=0$, if necessary. 
The entropy changes from zero for permutation matrices to $\ln N$ for the
uniform matrix $B_{\ast}$.

Due to the majorization of each column vector,
 the relation $C\prec B$ implies $H(B)\leq H(C)$, but 
 the converse is not true.
To characterise quantitatively the effect of entropy increase under the action
 of a bistochastic matrix $B$, let us define the {\textsl{weighted entropy}}
of matrix $B$ with respect to a probability vector
$\vec{y}=\left(y_{1},\ldots,y_{N}\right) $:

\begin{equation}
H_{\vec{y}}(B):=\sum_{k=1}^{N}y_{k}H(B_{k})=-\sum_{k=1}^{N}%
y_{k}\sum_{j=1}^{N}B_{jk}\ln B_{jk} \text{ ,}
\label{entwei}%
\end{equation}
where $B_k$ is a probability vector defined by $B_{k}:=\left(B_{1k},\ldots,B_{Nk}\right)$ 
for $k=1,\ldots,N$. In this notation $H(B) = H_{e_*}(B)$, where $e_* =
(1/N,\ldots,1/N)$. The weighted entropy allows one to write down the following
bounds for $H(B\vec{y})$

\begin{equation}
\max\left\{  H(\vec{y}),H_{\vec{y}}(B)\right\}  \leq
H(B\vec{y})\leq H(\vec{y})+H_{\vec{y}%
}(B)\text{ .}\label{ineq2}%
\end{equation}
the proof of which is provided elsewhere \cite{Sl01}. These bounds have certain
 implications in quantum mechanics. For example, if a non-unitary evolution
of the density operator under the action of random external field is
considered \cite{Ni99b}, they tell us how much the von Neumann entropy of the
mixed state may grow during each iteration.

Using the above proposition we shall show that the entropy of bistochastic matrices 
is subadditive. Namely, the following theorem holds:

\medskip

{\bf Theorem 2.} {\sl Let $A$ and $B$ be two bistochastic matrices. Then}
\begin{equation}
\max\left\{  H(A),H(B)\right\}  \leq H(AB)\leq H(A)+H(B)\text{ ,}
\label{ineq3}
\end{equation}
{\sl and, analogously,}
\begin{equation}
\max\left\{  H(A),H(B)\right\}  \leq H(BA)\leq H(A)+H(B)\text{ .}%
\label{ineq4}
\end{equation}
%\end{theorem}

%\begin{proof}
{\bf Proof.}
We put $C:=AB$ and consider stochastic vectors 
$\vec{y}_n :=\left(  c_{1n},\ldots,c_{Nn}\right)  $ for $n=1,\ldots,N$ (the
columns of the matrix $C$). Applying (\ref{ineq2}) we get  \[\max\left( 
H_{\vec{y}_{n}}(B),H\left( \vec{y}_n \right) \right) \leq H\left(
B\vec{y}_n\right) \leq H_{\vec{y}_n}(B)+H\left(  \vec{y}_n\right) .\]  Hence 
\[\max\left({\textstyle\sum\nolimits_{k=1}^{N}} c_{kn}%
{\textstyle\sum\nolimits_{j=1}^{N}} \eta\left( b_{jk}\right) ,
{\textstyle\sum\nolimits_{k=1}^{N}}
\eta\left( c_{kn}\right) \right) \leq
{\textstyle\sum\nolimits_{k=1}^{N}}
\eta\left(
{\textstyle\sum\nolimits_{j=1}^{N}}
b_{kj}c_{jn}\right)
\]
and
\[
{\textstyle\sum\nolimits_{k=1}^{N}}
\eta\left(
{\textstyle\sum\nolimits_{j=1}^{N}}
b_{kj}c_{jn}\right) \leq
{\textstyle\sum\nolimits_{k=1}^{N}}
c_{kn}%
{\textstyle\sum\nolimits_{j=1}^{N}}
\eta\left( b_{jk}\right) +
{\textstyle\sum\nolimits_{k=1}^{N}}
\eta\left( c_{kn}\right) \text{ ,}
\]
where $\eta\left( x\right) =-x\ln x$ for $x>0$. Multiplying the above 
equalities by $1/N$ and summing over $n=1,\ldots,N$ we get (\ref{ineq3}).
Setting $C=BA$ we obtain (\ref{ineq4}) in an analogous way.   $\Box$
%\end{proof}

\bigskip

\section{Ensembles of random unistochastic matrices}
\label{secIV}

\subsection{Unistochastic matrices}

To demonstrate that a given bistochastic matrix $B$ is unistochastic 
one needs to find unitary matrix $U$ such that $B_{ij}=|U_{ij}|^2$.
In other words one needs to find a solution of the
coupled set of nonlinear equations enforced by the unitarity
$UU^{\dagger}=U^{\dagger}U={\mathbb I}$,
\begin{equation}
 \sum_{j=1}^N \sqrt{B_{jk}B_{jl}} \exp(i \phi_{jk}- \phi_{jl}) = \delta_{kl}
 \ {\rm ~~for ~~ all ~~} 1\leq k<l\leq N
\label{unit1}
\end{equation}
for the unknown phases of each complex element 
$U_{jk}=\sqrt{B_{jk}} e^{i \phi_{jk}}$. 
The diagonal constraints for $k=l$ are fulfilled,
since $B$ is bistochastic. 

For $N=2$ all bistochastic matrices are orthostochastic
(see eg. Eq.(\ref{ttransform})),
 and so $\Omega_2^B=\Omega_2^U=\Omega_2^O$. This is not the case for higher
dimensions, for which there exist bistochastic matrices which are not
unistochastic, and so $\Omega_N^U \varsubsetneq \Omega_N^B$ for $N \geq 3$.
Thus $\Omega_N^U$ is not a convex set for $N \geq 3$, since it contains all
the permutation matrices and is smaller than their convex hull, which,
according to the Birkhoff theorem, is equal to $\Omega_N^B$ . Simple examples
of bistochastic matrices which are not unistochastic were already
provided (for $N=3$) by Schur and Hoffman \cite{MO79},

\begin{equation}
B_{1}=\frac{1}{2} \left[
\begin{array}{ccc}
1 & 1 & 0 \\
1 & 0 & 1 \\
0 & 1 & 1
\end{array}
\right] ,\text{ \ \ \ \ \ \ \ \ \ \ \ \ \ \ \ \ \ }B_{2}=\frac{1}{6} \left[
\begin{array}{ccc}
0 & 3 & 3 \\
3 & 1 & 2 \\
3 & 2 & 1
\end{array}
\right]    .
\label{matkontr} 
\end{equation}

%\vskip -2.0cm
\begin{figure} [htbp]
%\vskip -1.0cm
\begin{center}
\
\includegraphics[width=10.0cm,angle=0]{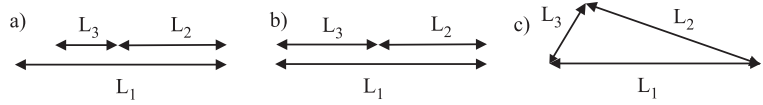}
\vskip 0.5cm
\caption{
Chain rule for unistochasticity for $N=3$: (see (\ref{necstouni1}) and
(\ref{necstouni2})),  a) the longest link $L_1 > L_2+L_3$ so the matrix
$B$ is not unistochastic, b) condition for orthostochasticity $L_1 = L_2+L_3$,
c) weaker condition for unistochasticity $L_1 \leq L_2+L_3$.}
 \label{fig2b} 
\end{center}
\end{figure}
%\vskip -1.0cm

\medskip

To see that there exists no corresponding unitary matrix $U$ 
we shall analyse constraints implied by the unitarity.
Define vectors containing square roots of the column (row) entries, e.g.,
${\vec v}_k:=\{ \sqrt{B_{k1}}, \sqrt{B_{k2}},..., \sqrt{B_{kN}} \}$.
The scalar products of any pair of any two different vectors
$\vec{v}_{k}\cdot \vec{v}_{l}$ consists of $N$ terms,
$L_n=\sqrt{ B_{kn}B_{ln}}$, $n=1,...,N$.
In the  case of $B_1$ from (\ref{matkontr}) 
the scalar product related to the two first columns
$\left(\vec{v}_{1},\vec{v}_{2}\right)$  consists of three terms 
$L_1=1$, $L_2=L_3=0$, which do not satisfy the triangle inequality.
 Thus it is not possible to find the corresponding
phases $\phi_{jk}$ satisfyig (\ref{unit1}).
This observation allows us to
obtain a set of {\sl necessary} conditions, a bistochastic $B$ must satisfy to
be unistochastic \cite{PZK01}:

\begin{equation}
\max_{m=1,\ldots,N} \sqrt{B_{mk}B_{ml}}~\le ~ \frac{1}{2}
 \sum_{j=1}^N \sqrt{B_{jk}B_{jl}} \ ,{\rm ~~for ~~ all ~~} 1\leq k<l\leq N
\label{necstouni1}
\end{equation}
and
\begin{equation}
\max_{m=1,\ldots,N} \sqrt{B_{km}B_{lm}}~\le ~ \frac{1}{2}
 \sum_{j=1}^N \sqrt{B_{kj}B_{lj}} \ , {\rm ~~for ~~ all ~~} 1\leq k<l\leq N \ .
\label{necstouni2}
\end{equation}

We shall refer to the above inequalities as to a `chain rule': the longest
link $L_1$ of a closed chain cannot be longer than the sum of all other links
$L_2+...+L_N$ - see Fig. \ref{fig2b}.
The set of $N(N-1)/2$ conditions (\ref{necstouni1}) (resp. (\ref{necstouni2}))
treats all possible pairs of the columns (resp. rows) of $B$. Only for
$N=3$ the both sets of conditions are equivalent
(since the equations (\ref{unit1}) for the phases may be separated \cite{Pa02}), 
but for $N\ge 4$ there exist matrices which 
satisfy only one class of the constraints.
For example, a bistochastic matrix  
\begin{equation}
B_4 = \frac{1}{100} \left[
\begin{array}{rrrr}
 $1$\ &  17 \  & 25 \ & 57  \\
 $38$\ &  38 \ & 24  \ & 0  \\
 $42$\ &  5 \  & 29 \ & 24  \\
 $19$\ &  40 \  & 22 \ & 19 
\end{array}
\right],  
\end{equation}
 found by Pako{\'n}ski \cite{Pa02}, 
 satisfies the row conditions (\ref{necstouni2}), but violates the
 column conditions (\ref{necstouni1}), so cannot be unistochastic
(the third term of the product of the roots of the first and the fourth column
 is larger than the sum of the remaining three terms). 
It is
easy to see that for an arbitrary $N$ these necessary conditions are satisfied
by any bistochastic $B$ sufficiently close to the van der Waerden matrix $B_*$,
for which all links of the chain are equal, $L_i=1/N$. It is then tempting to
expect that there exists an open vicinity of $B_*$ included in $\Omega_N^U$,
i.e., that $B_*$ lies in the interior of $\Omega_N^U$ and, consequently, that
$\Omega_N^U$ has positive volume. Certain conditions {\sl
sufficient} for unistochasticity were already found by Au-Yeung and Cheng
\cite{AYC91}, but they do not answer the question concerning the volume 
of $\Omega_N^U$. On the other hand, it is well known that the set of
unistochastic matrices is connected and compact \cite{He78}, and is not dense
in the set of bistochastic matrices \cite{Dj66}.

To analyse properties of bistochastic matrices it is convenient to introduce
 so called $T$-transforms, which in a sense reduce the problem to two
dimensions. The $T$-transform acts as the identity in all but two dimensions,
in which it has a common form of an orthostochastic matrix

\begin{equation}
{\tilde T}\left( \varphi \right) = \left[
\begin{array}{cc}
\cos^2 \varphi & \sin^2 \varphi \\
 \sin^2 \varphi & \cos^2 \varphi
\end{array}
\right]
{ \quad \quad \rm such \quad that} \quad \quad
{\tilde O}_2 \left( \varphi \right) = \left[
\begin{array}{cc}
\cos \varphi & \sin \varphi \\ - \sin \varphi & \cos \varphi
\end{array}
\right]
\label{ttransform}
\end{equation}
is orthogonal. 
Any matrix $B$ obtained as a sequence of at most $(N-1)$ $T-$transforms,
 $B = T_{N-1} \cdots T_2 T_1$,  where each $T_k$ acts in the
two-dimensional subspace spanned by the base vectors $k$ and $k+1$,
is orthostochastic. To show this it is enough to observe that 
each element of $B$ is a product of
non-trivial elements of the transformations $T_k$.
Hence taking its square root
and adjusting the signs one may find a corresponding orthogonal matrix defined
by the products of the elements of ${\tilde O}_k$ \cite{MO79,Bh97,Ni99}.
Although any product of an arbitrary number of T-transforms 
satisfies the chain--links conditions \cite{BE02}, 
for $N\ge 4$ there exist products of a finite number of
T-transforms (also called {\sl pinching matrices}) which are not 
unistochastic \cite{PT87}. In the same paper it is also shown that 
there exist unistochastic matrices which cannot be written 
as a product of $T$-transforms.

Consider a unistochastic matrix $B$ and the set ${\cal U}_B \subset U(N) $
 of all unitary matrices corresponding to $B$ in the sense that
$B_{ij}=|U_{ij}|^2$ for $i,j=1,\ldots,N$. Such sets endowed with 
appropriate probability measures play a role in the theory of quantum graphs
\cite{Ta00,PZK01,Ta01} and were called {\sl unitary stochastic ensembles} by
Tanner \cite{Ta01}. It is easy to see that these sets are invariant under the
operations $U \to V_{1}UV_{2}$, where $V_1$ and $V_2$ are arbitrary diagonal
unitary matrices. The dimensionality arguments suggest that, having $U \in
{\cal U}_B$ fixed, each element of ${\cal U}_B$ can be obtained in this way
\cite{false}. However, in general this conjecture is false and for certain
bistochastic matrices $B$ the set ${\cal U}_B$ is larger. This is shown in
Appendix \ref{sec:unibist}, in which a counterexample for $N=4$ is provided.

\subsection{Definition of ensembles}

To analyse random unistochastic matrices one needs to specify a probability 
measure in this set. As it was already discussed in the introduction,
unistochastic (USE) and orthostochastic (OSE) ensembles can be defined with
help of the Haar measure on the group of unitary matrices $U(N)$ and
orthogonal matrices $O(N)$, respectively \cite{Be01}. In other words the
bistochastic matrices

\begin{equation}
B^U_{ij}:=|U_{ij}|^2, \quad {\rm and} \quad B^O_{ij}:=(O_{ij})^2,
\label{uniort}
\end{equation}
pertain to USE and OSE respectively, if the matrices $U$ and $O$ are generated 
with respect to the Haar measures on the unitary (orthogonal) group.

Dynamical systems with time reversal symmetry are described by unitary symmetric 
matrices \cite{Haake}. The ensemble of these matrices, defined by $W=UU^T$,
is invariant with respect to orthogonal rotations, and is called {\sl circular
orthogonal ensemble} (COE). In an analogous way we may thus define the
following three ensembles of symmetric bistochastic matrices (SBM) 

\begin{equation}%
\begin{tabular}
[c]{llll}%
a)\qquad & $S_{1}:=BB^{T};$ & $\quad\mathrm{so}\quad$ & $S_{ij}:=\sum
_{k=1}^{N}|U_{ik}|^{2}|U_{jk}|^{2}$,$\medskip$\\
b)\qquad & $S_{2}:={\frac{1}{2}}(B+B^{T});$ & $\quad\mathrm{so}\quad$ &
$S_{ij}={\frac{1}{2}}(|U_{ij}|^{2}+|U_{ji}|^{2})$,$\medskip$\\
c)\qquad & $S_{3}:=|W_{ij}|^{2}=|(UU^{T})_{ij}|^{2};$ & $\quad\mathrm{so}%
\quad$ & $S_{ij}:=|\sum_{k=1}^{N}U_{ik}U_{jk}|^{2}, \medskip\medskip$
\end{tabular}
\label{bisym}
\end{equation}

where bistochastic matrices $B$ are generated according to USE 
(or, equivalently, unitary matrices $U$ are generated according to CUE).

\subsection{Spectra of random unistochastic matrices}

Since the sets of the bi-, uni--, and (ortho--)stochastic matrices are related by 
the inclusion relations: \linebreak $\Omega_N^O \subseteq \Omega_N^U \subseteq
\Omega_N^B \subset \Omega_N^S$, analogous relations must hold for the
supports of their spectra, $\Sigma_N^O \subseteq \Sigma_N^U \subseteq
\Sigma_N^B \subseteq \Sigma_N^S$. For $N=2$ the spectrum of a bistochastic
matrix must be real. The subleading eigenvalue $\lambda_2 = 2B_{11}-1$, which
allows one to obtain the distributions along the real axis. 
For USE the respective density is constant, $P_r(\lambda)=1/4$ for
$\lambda\in (-1,1)$, while for OSE it is given by the formula,
$P_r(\lambda)=1/(2\pi \sqrt{1-\lambda^2})$.
These densities are normalized to $1/2$, since for any
random matrix its leading eigenvalue is equal to unity.

For larger matrices we generated random unitary (orthogonal) matrices with respect to the Haar measure by means of the Hurwitz parameterisation \cite{Hu97} as described in \cite{ZK94,PZK98}. Squaring each element of the matrices
generated in this way we get random matrices typical for USE (OSE). In
general, the density of the spectrum of random uni--, (ortho--)stochastic matrices
may be split into three components:

a) two-dimensional density with the domain forming a subset $\Sigma_N^U$ ($\Sigma_N^O$)
 of the unit circle;

b) one-dimensional density at the real line described by the function $P_r(x)$ for $x\in [-1,1]$,

and

c) the Dirac delta $\frac{1}{N} \delta(z-1)$ describing the leading
eigenvalue,        which exists for every unistochastic matrix.

%\vskip -1.0cm
\begin{figure} [htbp]
%\vskip -1.0cm
   \begin{center}
\
 \includegraphics[width=10.0cm,angle=0]{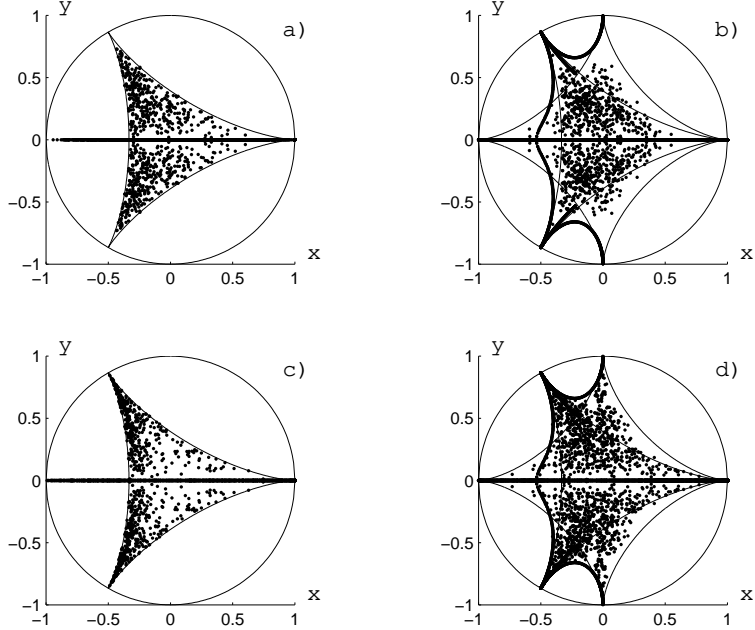}
\vskip 0.5cm
\caption{
Spectra of random unistochastic matrices of size a) $N=3$ ($1200$ matrices) 
and b) $N=4$ ($800$ matrices); spectra of random orthostochastic matrices of
size c) $N=3$ ($1200$ matrices) and d) $N=4$ ($800$ matrices). Thin lines
denote $3$-- and $4$--hypocycloids, while the thick lines represent the
$3$--$4$ interpolation arc  (part of it is shown in Fig.~2f). } \end{center}
 \end{figure}
%\vskip -2.0cm

%\medskip

Basing on the numerical results presented in Fig.~4a, we conjecture that
$\Sigma_3^U=\Sigma_3^O$ and consists of a real interval (already present for
$N=2$) and the inner part of the $3$--hypocycloid. This curve is drawn by a
point at a circle of radius $1/3$, which rolls (without sliding) inside the
circle of radius $1$. The parametric formula reads

\begin{equation}
\left\{
\begin{array}
[c]{ccc}
x & = & {\frac{1}{3}}(2\cos\phi+\cos2\phi),\medskip\\
\ y & = & {\frac{1}{3}}(2\sin\phi-\sin2\phi),
\end{array}
\right. \label{hyper3}
\end{equation}
where $\phi\in [0,2\pi)$.

To find the unistochastic matrices with spectra at the cycloid consider a 
two--parameter family of combinations of permutation matrices
$a^2{\mathbb I}+b^2P+c^2P^2$ with $a^2+b^2+c^2=1$. Here $P$ represents the 
nontrivial $3$ elements permutation matrix, $P=P_{(123)}$, so $P^3={\mathbb I}$. 
One--parameter family of these bistochastic matrices, which satisfy the
condition for unistochasticity, produce spectra located along the entire
hypocycloid. Consider the matrices

\begin{equation}
O_{3}(\varphi): =
 \left[
\begin{array}{ccc}
a & b & c \\
c & a & b \\
b & c & a
\end{array}
\right]
\quad \quad {\rm and} \quad \quad
 B_{3}(\varphi): =
 \left[
\begin{array}{ccc}
a^2 & b^2 & c^2 \\
c^2 & a^2 & b^2 \\
b^2 & c^2 & a^2
\end{array}
\right] ,
\label{hipo3}
\end{equation}
where their elements depend on a single parameter $\varphi \in [0,2\pi)$ and

\begin{equation}
\left\{ 
\begin{tabular}{lll}
$a=$ & $a(\varphi )=$ & $-\frac{1}{3}(1+2\cos \varphi ) \text{ ,}$ \\ 
$b=$ & $b(\varphi )=$ & $\frac{1}{3}(\cos \varphi -1)+\frac{1}{\sqrt{3}}\sin
\varphi \text{ ,}$ \\ 
$c=$ & $c(\varphi )=$ & $\frac{1}{3}(\cos \varphi -1)-\frac{1}{\sqrt{3}}\sin
\varphi \text{ .}$%
\end{tabular}
\right.   
\label{hyper3bi}
\end{equation}
It is easy to see that $a^2+b^2+c^2=1$ and $ab+bc+ca=0$, so $O_3$ is orthogonal 
and $B_3$ is orthostochastic. Simple algebra shows that the spectrum of
$B_3(\varphi)$ forms the $3$--hypocycloid given by (\ref{hyper3}) - see Appendix
\ref{sec:hypoc}.

An alternative approach, based on exponentiation of permutation matrices,
 leads to unistochastic matrices with spectrum on a hypocycloid. Let
$P_{N}:=P_{(12\cdots N)}$ be the nontrivial permutation matrix of size $N$.
Since $P_{N}$ is unitary, so is its power $(P_{N})^{\alpha}$. We define it 
for an arbitrary real exponent by 
 $(P_{N})^{\alpha}:=U^\dagger D^\alpha U$,
where $U$ is an unitary matrix diagonalizing $P_N$
and  $D$ represents the diagonal matrix of its eigenvalues --
 it is assumed that the eigenphases of such a matrix
belong to $[0,2\pi)$. Since $P_{N}^0={\mathbb I}_N$, one may
expect that defining the corresponding bistochastic matrices and varying the
exponent $\alpha$ from zero to unity one obtains an arc of a hypocycloid.
 This fact is true  as shown in Appendix \ref{sec:hypocun}, in which we 
prove a general result, valid for an arbitrary matrix size.

\medskip

{\bf Proposition 3.} {\sl
The spectra of the family of unistochastic matrices of size $N$}

\begin{equation}
B^{(N,\alpha)}_{ij}:= \bigr| \bigr(P_{N}^{\alpha}\bigl)_{ij}\bigl|^2
{\rm \quad  with  \quad} 
 \alpha \in  [ 0, N/ 2],
\label{hypenu}
\end{equation}
{\sl generate the $N$-hypocycloid (and
inner diagonal hypocycloids of the
radius ratio $k/N$ with $k=2,\ldots,N-1$).}

\begin{figure} [htbp]
\begin{center}
\
\includegraphics[width=12.0cm,angle=0]{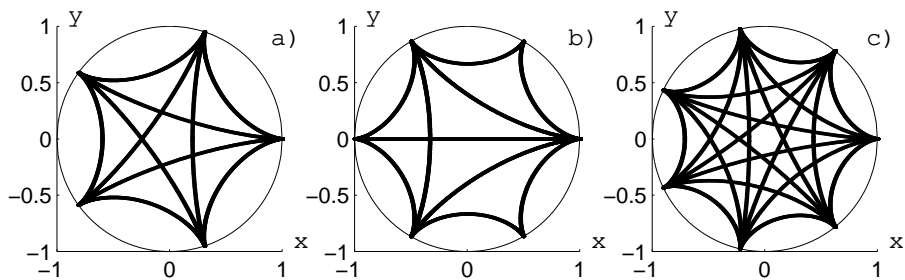}
\vskip 0.5cm
\caption{
Spectra of a family of bistochastic matrices
 $B^{(N,\alpha)}$ with $\alpha \in [0,N/2]$  obtained by exponentiation of
the permutation matrix $P_{N}$ plotted for a) $N=5$ (hypocycloids $5$ and
$5/2$),  b) $N=6$ (hypocycloids $6$, $3=6/2$ and $2=6/3$),  and c) $N=7$
(hypocycloids $7$, $7/2$ and $7/3$). } 
\end{center}
\end{figure}

Let $H_N$ denote the set bounded by $N-$hypocycloid. Fig.~4c suggests that
$H_3$ is contained in $\Sigma_3^O$. This conjecture may be supported by
considering the spectra interpolating between the origin $(0,0)$ and a
selected point on the hypocycloid. To find such a family we shall use the
Fourier matrix $F^{(N)}$ of size $N$ with elements $F^{(N)}_{kl}:=\exp(-2kl\pi
i/N)/\sqrt{N}$. Since amplitudes of all elements of this matrix are equal, the
corresponding bistochastic matrix equals to the van der Waerden matrix $B_*$
for which all subleading eigenvalues vanish. The matrix $P_{N}^{\alpha}$
generates a unistochastic matrix $B^{(N,\alpha)}$ with an eigenvalue at the
hypocycloid, so the family of unistochastic matrices related to
$\bigl(P_{N}^{\alpha}\bigr)^{\beta} (F^{(N)})^{1-\beta}$ provides an
interpolation between the origin and a selected point on the hypocycloid - see
Fig.~6. In other words, we conjecture that the spectra of a two parameter
family of unistochastic matrices obtained from $(P_{N})^{a} (F^{(N)})^{b}$
explore the entire $H_{N}$.

Numerical results obtained for $N=3$ random uni--, (ortho--)stochastic matrices allow us to
claim that there are no complex eigenvalues outside the 3-hypocycloid, so
that $\Sigma_3^U=\Sigma_3^O=H_2\cup H_3$. 
Interestingly, $H_3$ - the $3$-hypocycloid and its interior,
determines the set of all unistochastic matrices which 
belong to the cross section of $\Omega_3^B$ defined by the plane spanned by
$P_3, P_3^2$ and $P_3^3={\mathbb I}$ \cite{BE02}, see also Appendix
\ref{sec:hypoc}.

\begin{figure} [htbp]
\begin{center}
\
\includegraphics[width=10.0cm,angle=0]{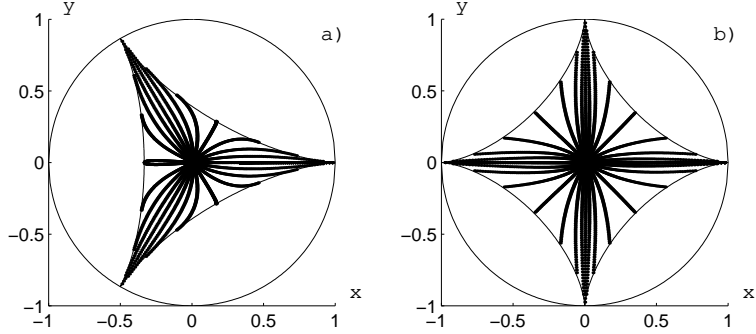}
\vskip 0.5cm
\caption{
Spectra of a family of bistochastic matrices obtained be squaring 
elements of unitary matrices $\bigl(P_{N}^{\alpha}\bigr)^{\beta}
(F^{(N)})^{1-\beta}$ with $\beta\in [0,1]$ and 
a) $N=3$, the curves are labelled by $\alpha=0,1/8,2/8,\ldots,3/2$, 
b) $N=4$, $\alpha=0,1/8,2/8,\ldots,2$.
For reference we plotted the hypocycloids with a thin line.
}
\end{center}
\end{figure}

Numerical results for spectra of random uni--, (ortho--)stochastic matrices for $N=4$ are shown 
in Fig.~4b~and~4d. The set $\Sigma_4^U$ includes the entire set $\Sigma_3^U$,
but also the $4$--hypocycloid, sometimes  called {\sl asteroid}, and  formed
by the solutions of the equation
\begin{equation}
x^{2/3} + y^{2/3} =1.
\label{asteroid}
\end{equation}

More generally, the spectra of orthostochastic matrices of size $N$ contain
 the $N$--hypocycloid $H_N$ 

\begin{equation}
\left\{
\begin{array}{ccc}
x&=&{\frac{1 }{N}} [(N-1) \cos \phi +\cos (N-1) \phi],  \\
y&=&{\frac{1 }{N}} [(N-1) \sin \phi -\sin (N-1) \phi],
\end{array}
\right.
\label{hyperN}
\end{equation}
with corner at $1$, as it is proved in Appendix \ref{sec:hypoc}. The set $\Sigma^U_N$ contains of 
the sets bounded by hypocycloids of a smaller dimension, which we denote by
$G_N: = \bigcup_{k=2}^N H_k$. 

However, as seen in Fig.~4b there exist some eigenvalues of unistochastic or 
even orthostochastic matrices outside this set.
In fact one needs to find interpolations between roots of unity of different
orders (the corners of $k$ and $n$--hypocycloid) based on the families of
orthostochastic matrices. Such a family interpolating between the corners of
$H_3$ and $H_4$ is plotted in Fig.~2f, since these bistochastic matrices are
orthostochastic. A general scheme of finding the required interpolations is
based on the fact that, if $O(\varphi)$ is a family orthogonal matrices and
$B_{ij}(\varphi)=O_{ij}^2(\varphi)$ is the corresponding family of
orthostochastic matrices, than another such  family is produced by the
 multiplication by some permutation matrix $P$: $O'(\varphi) = O(\varphi)P$.
For example, if the matrix $O_4(\varphi)$ of size $4$ contains elements
$O_{11}=O_{22}=1$ and the block diagonal matrix $O_2(\varphi)$ (see
(\ref{ttransform})), then the squared elements of the orthogonal matrices
$O_4(\varphi)P_{(1234)}$ provide orthostochastic matrices, the spectra of
which form the $3$--$4$ interpolating curve contained in $E_4$ and plotted in
Fig.~2f~and~4d. To obtain orthostochastic matrices, the spectra of which
provide the curve which joins both complex corners of $H_3$ and does not
belong to it (see Fig 4d), one should take the orthogonal matrix $O_{3,1}$ of
size $4$ with $O_{11}=1$ which contains the block diagonal matrix
$O_3(\varphi)$ (\ref{hipo3}), and create bistochastic matrices out of
$O_{3,1}(\varphi)P_{(12)(34)}$. The exact form of these families of
orthostochastic matrices is provided in Appendix \ref{sec:hypoc2}.

Let ${\tilde G}_N$ denote the set $G_N$ extended by adding the regions bounded
by the interpolations between all neighbouring corners of $H_k$ and
$H_n$ with $k<n \le N$ constructed in an analogous way as for $N=4$.
The sequence of hypocycloids with the neighbouring corners
is given by the order of fractions present in the
{\sl Farey sequence} and {\sl Farey tree} \cite{Schuster}. 
Our investigation of the support of the spectra
of unistochastic and orthostochastic matrices may be concluded by a relation
analogous  to (\ref{sigmabisto}) 

\begin{equation}
\bigcup_{k=2}^N H_k :=G_N \subset \Sigma_N^O \simeq \Sigma_N^U
\simeq {\tilde G}_N \subset E_N=:\bigcup_{k=2}^N Z_k,
\label{sigmaunisto}
\end{equation}
where the sign $\simeq$ represents the conjectured equality. Numerical results 
suggest that the support  $\Sigma_N$ is the same for both ensembles: USE and
OSE. However, the density of eigenvalues $P(z)$ is different for uni-- and
(ortho--)stochastic ensembles. The repulsion of the complex eigenvalues from the
real  line is more pronounced for the orthostochastic ensemble, as
demonstrated in Fig~4b~and~4d.

The larger $N$, the better the domains $\Sigma_N^O$ and $\Sigma_N^U$ fills
 the unit disk. On the other hand, the eigenvalues of a large modulus are
unlikely; the density $P(z)$ is concentrated in a close vicinity of the origin,
$z=0$. Moreover, in this case the weight of the singular part of the density
at the real line, decreases with the matrix size $N$. To characterise the
spectrum qualitatively we analysed the densities of the distributions $P_k(r)$
of the moduli of the largest eigenvalues ${\lambda}_k$.  The results obtained
for random unistochastic matrices are shown in Fig.~7a. Fig.~7b shows
analogous data for the singular values $\sigma_i$ of $B$ -- per definition
square roots of the real eigenvalues of the symmetric matrices $BB^T$
\cite{HJ98}. Since the singular values bound moduli of eigenvalues from above
\cite{HJ98}, the distribution $P(\sigma)$ is localised at larger values than
$P(r)$. Thus the expectation values satisfy  $\langle r_k\rangle <
\langle\sigma_k\rangle$ as shown in Fig.~7c~and~7d. The modulus of the second
eigenvalue decreases with matrix size as $N^{-1/2}$. These results are
consistent with the recent work of Berkolaiko \cite{Be01}, who suggested
describing the distribution $P_2(r)$ by the generalised extreme value
distribution \cite{LLR83}.

\vskip -3.9cm
\begin{figure} [htbp]
\begin{center}
\
\includegraphics[width=12.0cm,angle=0]{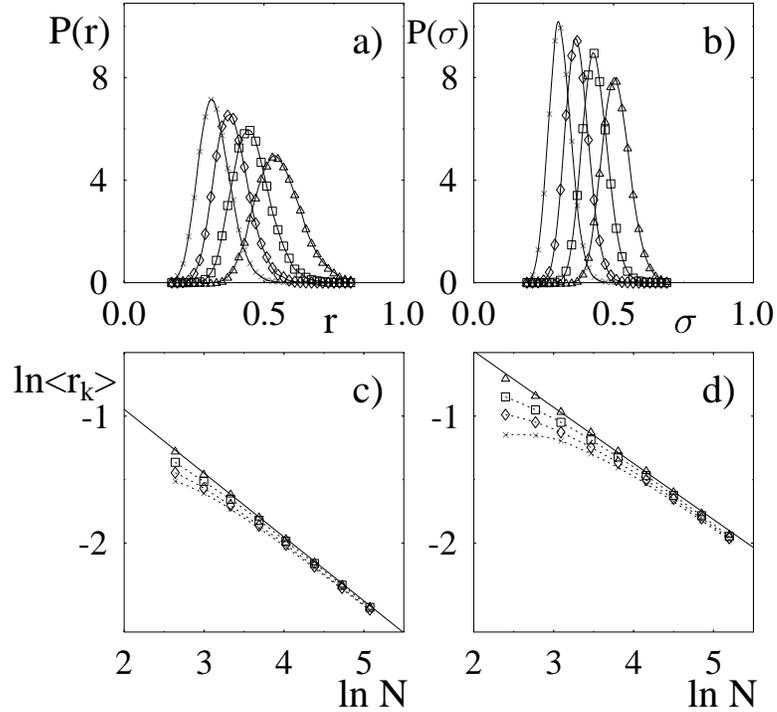}
\vskip  -2.9cm
\caption{
Unistochastic ensemble.  Distribution of the modulus of the 
second $(\triangle)$, third $(\square)$, fourth $(\Diamond)$ and fifth
$(\times)$ largest eigenvalues a), and singular value b) (lines are drawn to
guide the eye). Expectation value $\langle r_2\rangle$ for $k=2,3,4,5$ of the
subleading c) eigenvalues, and d) singular values as a function of the matrix
size $N$. Exponential fit, represented by solid lines, give $\langle
r_2\rangle \approx \exp(-0.503 )$ and  $\langle \sigma_2 \rangle \approx
\exp(-0.446)$ }.
\end{center} 
\end{figure}

\subsection{Average entropy}

We can compute the mean entropy averaged over the ensembles of unistochastic or 
orthostochastic matrices. Clearly we have

\begin{equation}
\left\langle H\right\rangle _{USE}=N\cdot\left\langle -\left|  U_{ij}\right|
^{2}\ln\left|  U_{ij}\right|  ^{2}\right\rangle _{CUE}
\end{equation}
and

\begin{equation}
\left\langle H\right\rangle _{OSE}=N\cdot\left\langle -\left|  O_{ij}\right|
^{2}\ln\left|  O_{ij}\right|  ^{2}\right\rangle _{COE} \text{ ,}
\end{equation}
where $U_{ij}$ and $O_{ij}$ are entries of unitary or orthogonal matrices, respectively. 
The exact formulae for the above averages were obtained by Jones in
\cite{Jo90,Jo91}. Using these results we get

\begin{equation}
\left\langle H\right\rangle _{USE}=\Psi(N+1)-\Psi(2)=\sum_{k=2}^{N}\frac{1}%
{k}\label{shauni}%
\end{equation}
and

\begin{equation}
\left\langle H\right\rangle _{OSE} = \Psi(N/2+1)-\Psi(3/2)=\left\{
\begin{tabular}
[c]{ll}%
$2\ln2-2+\sum\nolimits_{l=1}^{k}\frac{1}{l}$ & \qquad $N=2k$\\
$2\sum\nolimits_{l=1}^{k}\frac{1}{2l+1}$ & \qquad $N=2k+1$%
\end{tabular}
\right.  \text{ .}
\label{shaorto}%
\end{equation}
where $\Psi$ denotes the digamma function. For large $N$ the digamma function behaves logarithmically, $\Psi(N)\sim\ln N$, and both averages display similar asymptotic behaviour $\left\langle H\right\rangle _{USE}\approx\ln N-1+\gamma$ and $\left\langle H\right\rangle _{OSE}\approx\ln N-2+\gamma+\ln2\text{
,}$ where $\gamma\approx0.577$ stands for the Euler constant. Note that both averages 
are close to the maximal value $\ln N$, attained for orthostochastic matrices
corresponding to $B_{\ast}$, and their difference $\left\langle H\right\rangle
_{USE}-\left\langle H\right\rangle _{OSE}$ converges to $1-\ln2\approx0.307$.

\subsection{Average traces}

Traces of consecutive powers of bistochastic matrices, ${\rm tr}B^n$, are
quantities important in applications related to {\sl quantum chaos} \cite{KS97}.
In the following we evaluate the traces 
$t_{N,n}:= \langle {\rm tr}B^n \rangle_{\rm USE}$, averaged over the
unistochastic ensemble.

The spectrum of a bistochastic matrix size $2$ is real and may be written as 
$\operatorname*{Sp}(B)=\{1,y\}$ with $y \in [-1,1]$. For the unistochastic
ensemble $y=\cos 2\theta$, where the angle $\theta$ is distributed uniformly,
$P(\theta )= 2/ \pi$ for $ \theta \in [ 0, \pi /2]$. The average traces may be
readily expressed by the moments
 $\langle y^{k}\rangle$ for $k \in \mathbb{N}$, namely,

\begin{eqnarray}
t_{2,2k+1}  &=&1+\langle y^{2k+1}\rangle =1 \text{ ,}
\nonumber \\
t_{2,2k}     &=&
1+\langle y^{2k}\rangle =\frac{2k+2}{2k+1} \text{ .}
\label{tmk}
\end{eqnarray}

Since the average size of all the diagonal elements must be equal, 
$\langle |U_{ii}|^{2}\rangle=1/N$, so the average trace $ t_{N,1} =\langle
\sum_{i=1}^{N}|U_{ii}|^{2}\rangle =1$ and does not depend on the matrix size.
Using the results of Mello \cite{Me90}, who computed several averages over the
Haar measure on the unitary group $U(N)$, we derive in Appendix
\ref{sec:traces} the following formulae

\begin{equation}
t_{N,2} = 1+\frac{1}{N+1},
~~~~~~ {\rm and } ~~~~~~
t_{N,3} =
\left\{
\begin{array}{ccc}
 1                             ~~ & ~~{\rm for}~~ & ~~ N=2 \ ,  \\
 1+\frac{2}{N^{2}+3N+2} ~~ & ~~{\rm for} ~~ &~~ N \geq 3\ .
\end{array}
\right.
\label{tN23}
\end{equation}
Analogous expressions for larger $N$ may be explicitly written down as functions
of the Mello averages, but it is not simple to put them into a transparent
form. Numerical results support a conjecture that for arbitrary $N$ the
average traces tend fast to unity and the difference $t_{N,n}-1$ behaves as
$N^{1-n}$. This fact is related to the properties of the spectra discussed
above: for large $N$ the spectrum is concentrated close to the center of the
unit circle, so the contribution of all the subleading eigenvalues to the
traces becomes negligible. Related results on average traces of the symmetric
unistochastic matrices $BB^T$ are provided in \cite{Be01}.

\bigskip

\section{Closing Remarks}

In this paper we define the entropy for bistochastic matrices and proved its subadditivity. We also analysed special classes of bistochastic matrices: the unistochastic and orthostochastic matrices, and introduced probability measures on these sets. We found a characterization of complex spectra of unistochastic matrices in the unit circle and discussed the size of their subleading eigenvalue, which determines the speed of the decay of correlation in the dynamical system described by the matrices under consideration. Unistochastic matrices find diverse applications in different branches of physics and it is legitimate to ask, how a physical system behaves, if it is described by a random
unistochastic matrix \cite{Be01}. Since we succeeded in computation of mean values of some quantities (traces, entropy) averaged over the ensemble of
unistochastic matrices, our results provide some information concerning this issue.

On the other hand, several problems concerning bi--, uni--, and (ortho--)stochastic 
matrices remain open and we conclude this paper listing some of them:

a) Prove the conjecture that the union of the spectra of bistochastic matrices 
$\Sigma_N^B$ is equal to the sum of regular polygons $E_N$, i.e.,
$\Sigma_N^B= E_N $.

b) Prove that the support $\Sigma_3^U$ of spectra of unistochastic matrices of size 
$N=3$ contains exactly the interval $[-1,1]$ and the set bounded by the
$3$--hypocycloid $H_3$, i.e., $\Sigma_3^U=G_3$.

c) Find an analytical form of the boundaries of the set ${\tilde G}_N$ obtained of 
interpolations between the neighbouring corners of $k$--hypocycloids with
$k\le N$.

d) Show that the support $\Sigma_N^U$ of the spectra of unistochastic matrices contains
 ${\tilde G}_N$ and check whether both sets are equal, i.e., $\Sigma_N^U =
{\tilde G}_N$. 

e) Show that the supports of spectra of uni-- and (ortho--)stochastic matrices
are     the same, i.e., $\Sigma_N^U=\Sigma_N^O$.

f) Calculate probability distribution $P_N(z)$ of complex
 eigenvalues ensembles for uni--(ortho--)stochastic matrices.

g) Compute the expectation value of the subleading
   eigenvalue $\langle r_2\rangle$, averaged over uni--(ortho--)stochastic
ensemble.

h) Calculate averages over the unistochastic ensemble of other quantities
   characterizing ergodicity of stochastic matrices, including ergodicity
coefficients analysed by Seneta \cite{Se93} and entropy contraction
coefficient  introduced by Cohen et al. \cite{CIRRSZ93}.

i) Find necessary and sufficient conditions for a bistochastic matrix to be unistochastic.

j) Provide a full characterization of the set of all unitary matrices $U$,
    which lead to the same unistochastic matrix $B$, i.e., $B_{ij}=|U_{ij}|^2$
  for $i,j=1,\dots,k$.

\bigskip

This paper is devoted to the memory of late Marcin Po{\'z}niak,
with whom we enjoyed numerous fruitful discussions
on the properties of bistochastic matrices several years ago.
We are thankful to Prot Pako{\'n}ski for a fruitful interaction 
and also acknowledge helpful remarks of I.~Bengtsson, 
 G.~Berkolaiko,  G.~Tanner, and M.~Wojtkowski.
 Financial support by Komitet Bada{\'n}
Naukowych under the grant 2P03B-072~19 and
the Sonderforschungsbereich `Unordung und grosse Fluktuationen'
der Deutschen Forschungsgemeinschaft is gratefully acknowledged.

\appendix
\section{Unistochastic matrices stemming from a given bistochastic matrix}
\label{sec:unibist}

Let us consider two unitary $N\times N$ matrices $U$ and $W$ such that for all
$i,j=1,\ldots,N$,

\begin{equation}\label{uw}
|U_{ij}|^2=|W_{ij}|^2 \text{ ,}
\end{equation}
i.e., that the corresponding unistochastic matrices are the same. It is obvious that this happens if $W=V_1UV_2$ with $V_1$ and $V_2$ unitary diagonal. However the converse statement, i.e., that (\ref{uw}) implies existence of two diagonal unitary matrices $V_1$ and $V_2$ such that $U=V_1WV_2$, is false \cite{false}.  The plausibility of such conjecture is based on the following dimensional argument: we have $N^2$ real numbers $u_{ij}:=|U_{ij}|$ fulfilling $2N-1$ relations stemming from the normalisation of the rows:  

\begin{equation} \sum_{j=1}^N u_{ij}^2=1, \quad i=1,\ldots,N \text{ ,}
\label{rows}
\end{equation}
and the columns

\begin{equation}
\sum_{i=1}^N u_{ij}^2=1, \quad j=1,\ldots,N
\label{cols}
\end{equation}
of the unitary matrix $U$. The number of independent relations is less by one than the total number of equations in (\ref{rows}) and (\ref{cols}) since summing all equations in (\ref{rows}) over $i$ gives the same as summing all 
equations in (\ref{cols}) over $j$, namely $N=\rm{Tr}(U^{\dagger}U)$. 
On the other hand the left and right multiplications by unitary 
diagonal $V_1$ and $V_2$ introduce exactly $2N-1$ parameters (here the number
of the independent  parameters is diminished by one from the number of
non-zero elements of both  $D_1$ and $D_2$ since in the resulting matrix only
the differences of eigenphases  of $D_1$ and $D_2$ appear, so we can always
put one of the eigenphases of $D_1$, say,  to zero without changing the result
of the transformation $W\mapsto V_1WV_2$). The simplest counterexample we know involves the following unitary matrices $U$ and $W$

\begin{equation}
U = \frac{1}{2} \left[
\begin{array}{rrrr}
1 &  1 &  1           & 1            \\
1 & -1 & -e^{i\alpha} & e^{i\alpha}  \\
1 & -1 &  e^{i\alpha} & -e^{i\alpha} \\
1 &  1 & -1           & -1
\end{array}
\right], \quad
 W = \frac{1}{2} \left[
\begin{array}{rrrr}
1 &  1          &  1          &  1  \\
1 & -1          & -1          &  1  \\
1 & -e^{i\beta} &  e^{i\beta} & -1  \\
1 &  e^{i\beta} & -e^{i\beta} & -1
\end{array}
\right] \text{ ,}
\end{equation}
with $|U_{ij}|^2=|W_{ij}|^2=1/4$. It is a matter of simple explicit
calculations to show that there are no unitary diagonal $V_1$
and $V_2$ fulfilling $W=V_1 U V_2$,
if $\alpha, \beta \in [0,2\pi]$ and $\alpha\beta \ne 0$.

\section{Orthostochastic matrices with spectrum at hypocycloids}
\label{sec:hypoc}

In this appendix we construct orthostochastic matrices with spectra on
$N$-hypocycloids. Consider the following $N\times N$ permutation matrix

\begin{equation}
P:=\left[
\begin{array}{cccccc}
   0    &   1    &   0    & \ldots &   0    &   0    \\
   0    &   0    &   1    & \ldots &   0    &   0    \\
 \vdots & \vdots & \vdots &        & \vdots & \vdots \\
   0    &   0    &   0    & \ldots &   0    &   1    \\
   1    &   0    &   0    & \ldots &   0    &   0    \\
\end{array}
\right]
\text{ ,}
\label{P}
\end{equation}
where for simplicity the dimensionality index $N$ has been omitted.
 We have $P^N = \mathbb{I}$, $P^K \ne \mathbb{I}$ for $K<N$ and the
eigenvalues of $P$ equal $\exp\left(\frac{2\pi i}{N}k\right)$,
$k=0,1,\ldots,N-1$.

We shall discuss separately the cases of odd and even $N$.

First let $N=2K+1$ and

\begin{equation}\label{OB}
O:=\sum_{j=0}^{N-1}a_jP^j,\quad B:=\sum_{j=0}^{N-1}a_j^2P^j  .
\end{equation}
Observe that, since $P^m$ and $P^n$ do not have common non-zero entries for $m\ne n$, $0\le m,n\le N-1$, the elements of $B$ are squares of the corresponding elements of $O$. The eigenvalues of $O$ and $B$ read, respectively,

\begin{eqnarray}
  \Lambda_k&=&\sum_{j=0}^{N-1}a_j\exp\left(\frac{2\pi i}{N}kj\right),\quad
  k=0,1,\ldots,N-1 \label{eigO} \\
  \lambda_k&=&\sum_{j=0}^{N-1}a_j^2\exp\left(\frac{2\pi i}{N}kj\right),\quad
  k=0,1,\ldots,N-1.
  \label{eigB}
\end{eqnarray}
Inverting the discrete Fourier transforms in (\ref{eigO}) we obtain
\begin{equation}\label{aj2}
   a_j=\frac{1}{N}\sum_{k=0}^{N-1}\Lambda_k\exp\left(-\frac{2\pi  
i}{N}kj\right),\quad
    k=0,1,\ldots,N-1,
\end{equation}
which, upon substituting to (\ref{eigB}), gives the eigenvalues of $B$ in terms of the eigenvalues of $O$

\begin{eqnarray}\label{main1}
  \lambda_k=\frac{1}{N^2}\sum_{j=0}^{N-1}\sum_{l=0}^{N-1}\sum_{r=0}^{N-1}\Lambda_l
  \Lambda_r\exp\left(-\frac{2\pi i}{N}(l+r-k)j\right)=
  \frac{1}{N}\sum_{l=0}^{N-1}\sum_{r=0}^{N-1}\Lambda_l\Lambda_r\delta_{r,k-l}=
  \frac{1}{N}\sum_{l=0}^{N-1}\Lambda_l\Lambda_{k-l},
\end{eqnarray}
where the indices are counted modulo $N$.

Our aim is now to find a family of orthogonal matrices $O(\phi)$ such 
that when $\phi$ changes (from $0$ to $2\pi$, say) the eigenvalue
$\lambda_0(\phi)$ renders the $N$-hypocycloid $H_N$ in the complex plane, i.e.

\begin{equation}\label{lambad0}
  \lambda_0(\phi)=\frac{1}{N}\left[(N-1)e^{i\phi}+e^{-i(N-1)\phi}\right] .
\end{equation}
First observe that the desired result is achieved if

\begin{equation}\label{equi}
  \Lambda_k=\exp[i(k-K)\phi], \quad k=0,1,\ldots,2K=N-1.
\end{equation}
Indeed

\begin{eqnarray}
\lambda_0&=&\frac{1}{N}\sum_{l=0}^{N-1}\Lambda_l\Lambda_{-l}
=\frac{1}{N}\left(\Lambda_0^2+\sum_{l=1}^{N-1}\Lambda_l\Lambda_{-l}\right)
=\frac{1}{N}\left(\Lambda_0^2+\sum_{l=1}^{N-1}\Lambda_l\Lambda_{N-l}\right)
\nonumber \\
&=&\frac{1}{N}\left(e^{-2iK\phi}+\sum_{l=1}^{N-1}e^{i(l-K)\phi}e^{(N-l-K)\phi}\right)
=\frac{1}{N}\left(e^{-i(N-1)\phi}+(N-1)e^{i(N-2K)\phi}\right)
\\
&=&\frac{1}{N}\left[(N-1)e^{i\phi}+e^{-i(N-1)\phi}\right].
\end{eqnarray}

In order to construct an orthogonal matrix $O$ with the spectrum (\ref{equi}) it is enough to find an antisymmetric $A$ with the eigenvalues

\begin{equation}\label{equia}
  \mu_k=i(k-K),\quad k=0,1,\ldots,2K.
\end{equation}
If $A$ is a polynomial in $P$ then so is $O=\exp(A\phi)$, moreover $O$ is orthogonal and has the desired spectrum (\ref{equi}). Let us thus write

\begin{equation}\label{A}
A:=\sum_{j=1}^K \alpha_j\left(P^j-P^{N-j}\right),
\end{equation}
which is clearly antisymmetric, with the eigenvalues

\begin{equation}\label{eigA}
  \mu_k=\sum_{j=1}^K \alpha_j\left[\exp\left(\frac{2\pi i}{N}jk\right)
  -\exp\left(\frac{2\pi i}{N}(N-j)k\right)\right]=
  2i\sum_{j=1}^K \alpha_j\sin\left(\frac{2\pi}{2K+1}kj\right),
  \quad k=0,1,\ldots,2K.
\end{equation}
Obviously $\mu_0=0$ and $\mu_{2K+1-k}=\mu_k$, $k=1,2,\ldots,K$. To fulfil (\ref{equia}) it is thus enough that

\begin{equation}\label{rnie}
  2\sum_{j=1}^K \alpha_j\sin\left(\frac{2\pi}{2K+1}kj\right)=k,\quad  
k=1,2,\ldots,K.
\end{equation}
Using

\begin{equation}\label{ortsin}
\sum_{j=1}^K \sin\left(\frac{2\pi}{2K+1}kj\right)
\sin\left(\frac{2\pi}{2K+1}mj\right)
=\frac{2K+1}{4}\,\delta_{mk},
\end{equation}
we solve (\ref{rnie}) for $\alpha_j$,

\begin{equation}
  \alpha_j=\frac{2}{2K+1}\sum_{k=1}^Kk\sin\left(\frac{2\pi}{2K+1}kj\right) .
\label{aj2k1}
\end{equation}

For $N$ even, $N=2K$, the construction is very similar. We introduce $E:=\text{diag}(1,1,\ldots,1,-1)$ and define $\tilde{P}:=EP$ which has eigenvalues $\exp\left(\frac{2\pi i}{N}(k+\frac{1}{2})\right)$, $k=0,1,\ldots,N-1$, and construct the matrix $O$ as a polynomial in $\tilde{P}$ rather than in $P$

\begin{equation}\label{Otilde}
  \tilde{O}:=\sum_{j=0}^{N-1}a_j\tilde{P}^j ,
\end{equation}
with the eigenvalues:

\begin{equation}\label{Lambdatilde}
  \tilde\Lambda_k=\sum_{j=0}^{N-1}a_j\exp\left(\frac{2\pi i}{N}
  \left(k+\frac{1}{2}\right)j\right),\quad k=0,1,\ldots,N-1
\end{equation}
The matrix $B$ of the squared elements of $O$ is, as previously, the following polynomial in $P$,

\begin{equation}\label{Btilde}
  B:=\sum_{j=0}^{N-1}a_j^2P^j .
\end{equation}
Using exactly the same method as above we express the eigenvalues $\lambda_k$ of $B$ are given in terms of $\Lambda_j$

\begin{equation}\label{main2}
  \lambda_k=\frac{1}{N}\sum_{l=0}^{N-1}\tilde\Lambda_l\tilde\Lambda_{k-l+1}.
\end{equation}
Now if

\begin{equation}\label{equiA2}
\tilde\Lambda_k=\exp\left[i\left(k-K+\frac{1}{2}\right)\phi\right],
\quad k=0,1,\ldots,2K-1=N-1,
\end{equation}
then

\begin{eqnarray}\label{hypoeven}
\lambda_{N-1}&=&\frac{1}{N}\sum_{l=0}^{N-1}\tilde\Lambda_l\tilde\Lambda_{N-l}
=\frac{1}{N}\left(\tilde\Lambda_0^2+\sum_{l=1}^{N-1}
\tilde\Lambda_l\tilde\Lambda_{N-l}\right)
=\frac{1}{N}\left(e^{-i(2K-1)\phi}
+\sum_{l=1}^{N-1}e^{i(l-K+1/2)\phi}e^{(N-l-K+1/2)\phi}\right)
\nonumber \\
&=&\frac{1}{N}\left(e^{-i(N-1)\phi}+(N-1)e^{i(N-2K+1)\phi}\right)
=\frac{1}{N}\left[(N-1)e^{i\phi}+e^{-i(N-1)\phi}\right].
\end{eqnarray}
In full analogy with the case of odd $N$ we look for an antisymmetric matrix $\tilde{A}$ with the eigenvalues

\begin{equation}\label{eigA2}
  \tilde\mu_k=i\left(k-K+\frac{1}{2}\right),
\end{equation}
in the form of a polynomial in $\tilde{P}$

\begin{equation}\label{Atilde}
\tilde{A}
:=2^{1/2}\tilde\alpha_K\tilde P^K+\sum_{j=1}^{K}
\tilde\alpha_j\left(\tilde{P}^j+\tilde{P}^{N-j}\right).
\end{equation}
Since, as it is easy to check,  $(\tilde{P}^j)^T=-\tilde{P}^{N-j}$ for $j=1,2,\ldots,K$, the matrix $A$ is indeed antisymmetric for arbitrary real $\alpha_j$, $j=1,2,\ldots,K$. The eigenvalues of $A$ read:

\begin{equation}\label{eigAtilde}
\tilde{\mu}_k=2^{1/2}i(-1)^k\alpha_K+
2i\sum_{j=1}^{K-1}\tilde\alpha_j
\sin\left[\frac{\pi}{K}\left(k+\frac{1}{2}\right)j\right],\quad
k=0,1,\ldots,2K-1.
\end{equation}
Using arguments similar to those in the odd $N$ case, we conclude that the choice

\begin{eqnarray}\label{altilde}
\tilde\alpha_j&=&\frac{1}{K}\sum_{k=0}^{K-1}
\sin\left[\frac{\pi}{K}\left(k+\frac{1}{2}\right)j\right]\left(k-K+\frac{1}{2}\right),
\quad j=1,2,\ldots,K-1 \\
\tilde\alpha_K&=&-\frac{\sqrt{2}}{4},
\end{eqnarray}
leads to the desired result (\ref{eigA2}) and, consequently, the $2K$-hypocycloid (\ref{hypoeven}).

\section{Unistochastic matrices with spectrum at hypocycloids}
\label{sec:hypocun}

The above constructed matrices with spectra on $N$-hypocycloids were {\em orthostochastic}. 
If the desired matrix should be merely {\em unistochastic}, but not
necessarily orthostochastic, the construction is even simpler. To this end let
us consider the matrix

\begin{equation}\label{Palpha}
  P^\alpha:=U^\dagger D^\alpha U,
\end{equation}
where $U$ is an unitary matrix diagonalizing $P$, where $P$ is given by (\ref{P})), and $D$ is a diagonal matrix

\begin{equation}\label{D}
  D:=\mathrm{diag}\left(1,e^{2\pi i/N}, e^{4\pi i/N},\ldots,e^{2(N-1)\pi
  i/N}\right)
\end{equation}
with the eigenvalues of $P$ on the main diagonal. Hence, consequently

\begin{equation}\label{Dalpha}
  D^\alpha=\mathrm{diag}(\Lambda_0,\Lambda_1,\ldots,\Lambda_{N-1}),
\end{equation}
where $\Lambda_k:=\exp(2k\alpha\pi i/N)$, $k=0,1,\ldots,N-1$, are the eigenvalues of $P^\alpha$. Obviously $P^\alpha$ is unitary, and as a function of $P$ can be written in the form $P^\alpha:=\sum_{j=0}^{N-1}a_jP^j$, which gives for the eigenvalues

\begin{equation}\label{}
\Lambda_k=e^{2k\alpha\pi i/N}=\sum_{j=0}^{N-1}a_j\exp\left(\frac{2\pi
i}{N}kj\right),\quad
  k=0,1,\ldots,N-1.
\end{equation}
As previously we obtain the coefficients $a_j$ by inverting the discrete Fourier transform

\begin{equation}\label{aju}
 a_j=\frac{1}{N}\sum_{k=0}^{N-1}\Lambda_k\exp\left(-\frac{2\pi i}{N}kj\right),\quad
    k=0,1,\ldots,N-1.
\end{equation}
The associated unistochastic matrix reads thus

\begin{equation}\label{Balpha}
   B:=\sum_{j=0}^{N-1}|a_j|^2P^j.
\end{equation}
and has as the eigenvalues
\begin{equation}\label{lambdaku}
\lambda_k=\sum_{j=0}^{N-1}|a_j|^2\exp\left(\frac{2\pi i}{N}kj\right),\quad
  k=0,1,\ldots,N-1,
\end{equation}
i.e.

\begin{eqnarray}\label{main1u}
  \lambda_k=\frac{1}{N^2}\sum_{j=0}^{N-1}\sum_{l=0}^{N-1}\sum_{r=0}^{N-1}\Lambda_l
  \Lambda_r^\ast\exp\left(-\frac{2\pi i}{N}(l+r-k)j\right)=
  \frac{1}{N}\sum_{l=0}^{N-1}\sum_{r=0}^{N-1}\Lambda_l\Lambda_r^\ast\delta_{r,l-k}=
  \frac{1}{N}\sum_{l=0}^{N-1}\Lambda_l\Lambda_{l-k}^\ast,
\end{eqnarray}
hence

\begin{eqnarray}\label{lambda1u}
  \lambda_1=\frac{1}{N}\left(\Lambda_0\Lambda_{N-1}^\ast +  
\sum_{l=1}^{N-1}\Lambda_l\Lambda_{l-k}^\ast\right)=\frac{1}{N}\left(e^{-2(N-1)\alpha\pi
  i/N}+(N-1)e^{2\alpha\pi i/N}\right),
\end{eqnarray}
which renders the $N$-hypocycloid for $\alpha\in[0,N/2]$. Similarly,
 the further eigenvalues $\lambda_k$ generate the inner hypocycloids
(e.g. 3-hypocycloid for $N=6$ -- see Fig. 5), what
proves Proposition 3. 

Observe that for $N=3$ the permutation matrices
$P,P^2$ and $P^3={\mathbb I}_3$ form an equilateral
triangle (in sense of the Hilbert--Schmidt distance,
which is induced by the Frobenius norm of a matrix,
$||P||:=\sqrt{PP^{\dagger}}$). Comparing Eq. (C6) and (C7) with $k=1$ we see
that both quantities have the same structure and the same dependence
on the coefficients $a_j$, which are implicit functions of the
parameter $\alpha$. Therefore, varying this parameter 
we obtain the very same curves in two entirely different spaces:
 the eigenvalue $\lambda_1=\lambda_1(\alpha)$
 provides the $3$--hypocycloid in the plane of complex spectra, 
while the family of unistochastic matrices $B=B(\alpha)$ 
forms the same hypocycloid in the two--dimensional cross-section of the
four--dimensional body of $N=3$ bistochastic matrices determined by 
$P,P^2$ and $P^3$. 

\section{Interpolation between corners of two hypocycloids}
\label{sec:hypoc2}

In this appendix we provide a discuss a family of unistochastic 
 matrices of size $4$, the spectra of which are not contained in the sum of
$2$, $3$ and $4$--hypocycloids. To find an interpolation between the corners
of $3$ and $4$--hypocycloids consider the orthogonal matrix
$\tilde{O}_{3,4}(\varphi)=O_4(\varphi)P_{1234}$,  as defined in section
\ref{secIV},

\begin{equation}
\tilde{O}_{3,4}(\varphi) := 
\left[
\begin{array}{cccc} 
1 & 0 & 0 & 0 \\  
0 & 1 & 0 & 0 \\  
0 & 0 & \cos \varphi & \sin \varphi \\  
0 & 0 & -\sin \varphi & \cos \varphi 
\end{array} 
\right] \left[
\begin{array}{cccc} 
0 & 1 & 0 & 0 \\  
0 & 0 & 1 & 0 \\  
0 & 0 & 0 & 1 \\  
1 & 0 & 0 & 0 
\end{array} 
\right]
\allowbreak =\allowbreak  
\left[
\begin{array}{cccc} 
0 & 1 & 0 & 0 \\  
0 & 0 & \cos \varphi & \sin \varphi \\  
0 & 0 & -\sin \varphi & \cos \varphi \\  
1 & 0 & 0 & 0 
\end{array} 
\right] 
\label{inter34a}
\end{equation}
For $\varphi$ varying in $[0,\pi/2]$ this family interpolates between 
a four--elements permutation $P_{1234}$ and a matrix, the absolute values of
which represent a three--elements permutation $P_{124,3}$. Thus the spectra of
the corresponding bistochastic matrices give an interpolation between  the
third and the fourth roots of identity. As shown in Fig.~4 this interpolation
is located {\sl outside} the hypocycloids $H_3$ and $H_4$, so the support
$\Sigma_4^U$ is larger than their sum $G_4$. Repeating the argument with the
multiplicative interpolation between any such a matrix and the Fourier matrix
$F^{(4)}$ we conclude that all points {\sl inside} the set bounded by this
interpolation belong to the support  $\Sigma_4^U$. An analogous scheme allows
us to find an $(N-1)\longleftrightarrow N$ interpolation. For example, the
family of orthogonal matrices $\tilde{O}_{4,5}(\phi)=O_5(\phi)P_{12345}$ of
size $5$ gives an interpolation between $1^{1/5}$ and $i=1^{1/4}$. 

To find the missing $N=4$ interpolation for the negative real part of the
 eigenvalues consider a permutation of the orthogonal matrix which contains
the block $O_3$ responsible for the $3$--hypocycloid, 

\begin{equation}
\tilde{O}_{3,2}:=
 \left[
\begin{array}{cccc} 
1 & 0 & 0 & 0 \\  
0 & a & b & c \\  
0 & c & a & b \\  
0 & b & c & a 
\end{array} 
\right]  \left[
\begin{array}{cccc} 
0 & 1 & 0 & 0 \\  
1 & 0 & 0 & 0 \\  
0 & 0 & 0 & 1 \\  
0 & 0 & 1 & 0 
\end{array} 
\right]
\allowbreak =\allowbreak  
\left[
\begin{array}{cccc} 
0 & 1 & 0 & 0 \\  
a & 0 & c & b \\  
c & 0 & b & a \\  
b & 0 & a & c 
\end{array}  
\right],
 \label{inter34b}
\end{equation}
where its elements $a$, $b$, and $c$ are function of the angle $\varphi$ as given in (\ref{hyper3bi}). Then the spectra of the corresponding orthostochastic matrices, obtained by squaring the elements of the above orthogonal matrices, 

\begin{equation}
\tilde{B}_{3,4}(\varphi):=  
\left[
\begin{array}{cccc} 
0 & 1 & 0 & 0 \\  
0 & 0 & \cos ^{2}\varphi  & \sin ^{2}\varphi \\  
0 & 0 & \sin ^{2} \varphi  & \cos ^{2}\varphi  \\  
1 & 0 & 0 & 0 
\end{array} 
\right] \quad {\rm and} \quad 
 \tilde{B}_{3,2}(\varphi):=
  \left[
\begin{array}{cccc} 
0 & 1 & 0 & 0 \\  
a^{2} & 0 & c^{2} & b^{2} \\  
c^{2} & 0 & b^{2} & a^{2} \\  
b^{2} & 0 & a^{2} & c^{2} 
\end{array} 
\right] 
 \label{inter34c}
\end{equation}
provide the required interpolations located outside hypocycloids 
$H_3$ and $H_4$ (see Fig.~4b~and~4d). For larger dimensionality an analogous
construction has to be performed to get the interpolations between the
neighbouring roots of identity.

\section{Average traces of unistochastic matrices}
\label{sec:traces}

To evaluate the average traces  we rely on the results of Mello \cite{Me90}, who computed  various averages, $\langle . \rangle$, over the Haar measure on the unitary group $U(N)$. In particular, he found average values of the following quantities constructed of elements $U_{ab}$ of a unitary matrix of size $N$

\begin{equation}
Q_{b_1 \beta_1, \cdots ,b_m\beta_m}^{a_1 \alpha_1, \cdots ,a_k\alpha_k}
:=
\langle (U_{b_1\beta_1}\cdots U_{b_m\beta_m})
 (U_{a_1\alpha_1}\cdots U_{a_k\alpha_k})^{\ast }\rangle.
\label{Mell1}
\end{equation}
\bigskip
Mean trace of a squared unistochastic matrix, defined by $B_{ij}=|U_{ij}|^2$, reads

\begin{equation}
t_{N,2}:=\langle {\rm Tr} B^{2}\rangle_{USE}
=
\langle \sum_{i,k=1}^{N}(U_{ik}U_{ik}^{\ast })(U_{ki}U_{ki}^{\ast
}) \rangle_{U(N)}
=\sum_{i,k=1}^{N}Q_{ik,ki}^{ik,ki} =
NQ_{11,11}^{11,11}+N(N-1)Q_{12,21}^{12,21},
\end{equation}
since the symmetry of the problem allowed us to group together the terms according to the number of different indices. Using the results of Mello
$Q_{11,11}^{11,11}=2/(N(N+1))$ and $Q_{12,21}^{12,21}=1/(N(N+1))$ we get $t_{N,2}=(N+2)/(N+1)$.

The mean trace of $B^3$ reads

\begin{equation}
 t_{N,3} :=\langle {\rm Tr} B^{3}\rangle_{USE}=
NQ_{11,11,11}^{11,11,11}+3N(N-1)Q_{11,12,21}^{11,12,21}+
N(N-1)(N-2)Q_{12,23,31}^{12,23,31}.
\label{an3}
\end{equation}

The data in Mello's paper allow us to find
$Q_{11,11,11}^{11,11,11}=6/[N(N+1)(N+2)]$,
$Q_{11,12,21}^{11,12,21}=1/[(N+2)(N^2-1)]$,
and
$Q_{12,23,31}^{12,23,31}= (N^2-2)/[(N(N^2-1)(N^2-4)]$,
where the last term (with three different indices) is present only for $N\ge 3$.
Substituting these averages into (\ref{an3}) we arrive with the result (\ref{tN23}).

In the general case of arbitrary $n$ we may write a formula

\begin{equation}
t_{N,n} :=\langle {\rm Tr} B^{n}\rangle_{USE}
=\sum_{i_1,\dots i_n}^{N} Q^{i_1i_2,i_2i_3, \dots ,i_{n-1}i_n,i_n i_1}%
_{i_1i_2,i_2i_3, \dots ,i_{n-1}i_n, i_n i_1},
\label{annn}
\end{equation}
which is explicit, but not easy to simplify. An analogous computation for the ensemble of symmetric unistochastic matrices may be based on results of Brouwer and Beenakker \cite{BB96}, who computed the averages (\ref{Mell1}) for COE.

\end{document}